\documentclass[11pt]{article}
\usepackage[a4paper, total={6.5in, 9in}]{geometry}
\usepackage{amsmath}
\DeclareMathOperator{\Tr}{Tr}
\usepackage{amsfonts}
\usepackage{mathtools}
\usepackage{hyperref}
\usepackage{breqn} 
\usepackage{bm}
\usepackage{graphicx}
\usepackage{epstopdf}
\usepackage{subcaption}
\usepackage{authblk}
\usepackage{color,soul}
\usepackage{tikz-cd} 
\usepackage{comment}
\usepackage[toc]{appendix}
\DeclarePairedDelimiterX{\norm}[1]{\lVert}{\rVert}{#1}
\def\equationautorefname~#1\null{Equation (#1)\null}

\title{Stochastic Dynamical Systems Developed on Riemannian Manifolds}
\author[1]{Mariya Mamajiwala} 
\affil[1]{Department of Statistical Science, University College London, United Kingdom}
\author[,2,3]{Debasish Roy \thanks{Corresponding author; Email: royd@iisc.ac.in}}
\affil[2]{Centre of Excellence in Advanced Mechanics of Materials, Indian
Institute of Science, Bangalore 560012, India }
\affil[3]{Computational Mechanics Lab, Department of Civil Engineering, Indian
Institute of Science, Bangalore 560012, India}

\date{}

\begin{document}
\maketitle
\begin{abstract} 
We propose a method for developing the flows of stochastic dynamical systems, posed as Ito's stochastic differential equations, on a Riemannian manifold identified through a suitably constructed metric. The framework used for the stochastic development, viz. an orthonormal frame bundle that relates a vector on the tangent space of the manifold to its counterpart in the Euclidean space of the same dimension, is the same as that used for developing a standard Brownian motion on the manifold. Mainly drawing upon some aspects of the energetics so as to constrain the flow according to any known or prescribed conditions, we show how to expediently arrive at a suitable metric, thus briefly demonstrating the application of the method to a broad range of problems of general scientific interest. These include simulations of Brownian dynamics trapped in a potential well, a numerical integration scheme that reproduces the linear increase in the mean energy of conservative dynamical systems under additive noise and non-convex optimization. The simplicity of the method and the sharp contrast in its performance  vis-\'a-vis the correspondent Euclidean schemes in our numerical work provide a compelling evidence to its potential.
   
\vspace{10pt}
    \end{abstract} 
    
\section{Introduction}
As a tool in the modelling of physical phenomena, physicists have long grappled with non-Euclidean differential geometry -- starting perhaps with Einstein's work on the general theory of relativity based on a four dimensional Riemannian manifold \cite{einstein1914foundation} to the modelling of shells in the mechanics of solids \cite{kupferman2014riemannian}.  In the physical world, a Riemannian manifold affords an ideal framework to study evolutions on or of generally curved objects. Unlike the Euclidean space, two tangent spaces to any two distinct points on a curved Riemannian manifold are not canonically isomorophic \cite{ClellandJeanneN}. In order to move between two neighbouring points on such a manifold, the notion of a connection is therefore introduced which usurps the traditional concept of derivative (or differential) in the Euclidean setting. Once subsumed within a mathematical model, this concept therefore enables a precise tracking of the evolution of a field on a Riemannian manifold, i.e. a curved hypersurface. This construct is thus useful, not just in a more insightful modelling of the physical world, but in the development of substantively more robust numerical algorithms that are more informed of the intrinsic structure or the given constraints of a flow generated by the model at hand. As an example, one may consider the flow of a Hamiltonian dynamical system that evolves on a constant-energy hypersurface, which is a manifold embedded in the ambient phase space of positions and momenta. Unfortunately, despite some promising work done in the broad area of optimization, efforts at developing numerical schemes on Riemannian manifolds appear to be somewhat scarce.

One such well-established method for solving constrained optimization problems in a deterministic setting is the gradient projection method (see \cite{LuenbergerRef4} and \cite{LuenbergerRef5} discussing this for linear and non-linear constraints respectively). In the work of Luenberger \cite{Luenberger1972}, the point of view was to consider an analogue of the geodesic on a constrained hypersurface, which was considered a Riemannian manifold, within a constrained optimization problem. In that setting, the geodesic would be a straight line for an unconstrained problem. Exploiting the Riemannian connection, i.e. the notion of the covariant derivative, Smith \cite{smith1994}  has proposed extensions of Newton's method and the conjugate gradient method to a Riemannian manifold setting. A Riemannian manifold variant of the BFGS quasi-Newton scheme may be found in \cite{ring2012optimization}. The perspective we adopt in this work is however able to pose even an unconstrained optimization problem on a Riemannian manifold.

For non-convex, global optimization problems, stochastic search schemes are typically preferred over deterministic, gradient-based methods. Most stochastic algorithms for such global optimization are based on heuristics, common examples being particle swarm optimization \cite{KennedyPSO1995}, ant colony optimization \cite{dorigo2006ant}, genetic algorithms \cite{whitley1994genetic}, etc. These methods typically work without needing explicit information on the derivatives of the objective functional. Though widely used, they generally lack a rigorous mathematical basis. COMBEO \cite{combeo} is a stochastic algorithm that poses optimization as a martingale problem. Although it is based on a sound mathematical principle, its implementation lacks the simplicity and the intuitive features of methods based on metaheuristics. A much simpler variant of optimization based on stochastic search is provided by the family of stochastic approximation schemes \cite{robbinsMonro1951}. Using a retraction mapping which is an expedient means to implement the exponential map, \cite{shah2019stochastic} has recently extended stochastic approximation to the case where the constraint set constitutes a Riemannian manifold.     

The flow generated by a dynamical system, with or without any additionally specified constraints, typically has a structure of its own. For instance, a conservative dynamical system must evolve over a constant-energy hypersurface which could be non-Euclidean. In this work, we are specifically concerned with the posing and solutions for stochastic dynamical systems, governed by a set of stochastic differential equations (SDEs), in the Riemannian manifold setting. For this purpose, we exploit the principle of stochastic development \cite{HsuElton}, thus bypassing the generally problematic issue of embedding within an ambient Euclidean space. Having identified the manifold structure associated with a stochastic dynamical system based on a metric, we use an orthonormal frame bundle structure to relate a vector in the Euclidean space with its unique counterpart on a tangent space to the manifold of the same dimension. More interestingly, the motion governed by a vector field in the Euclidean space could be tracked on the Riemannian manifold through its projection via the horizontal part of the tangent space to the orthonormal frame bundle. This requires a parallel transport of vectors and hence information about the Levi-Civita connection on the manifold, which is also shared by the horizontal part of the tangent space to the frame. This is the essence of stochastic development, which we use to rewrite an SDE with non-zero drift on a Riemannian manifold. For purposes of illustration and to demonstrate the wide spread of possible applications of this approach, we consider the problems of Brownian motion restricted by a potential well, energy-drift conserving numerical integration of a noisy, undamped nonlinear mechanical oscillator without an external force, and global optimization involving a non-convex objective function. In each case, as appropriate to the nature of response we seek, the Riemannian metric and the associated connection are derived based on a chosen energy-like function. The metric and connection are in turn used to obtain the developed SDEs, which are numerically integrated with the basic Euler-Maruyama explicit method. In the case of the non-convex optimization problem, for instance, the objective function (which is strictly positive) is itself interpreted as the energy and the design variables are evolved according to a stochastically developed Langevin dynamics implemented along with a simulated annealing scheme. In order to emphasize the role played by stochastic development, we always contrast the solutions of developed SDEs with those in the classical Euclidean setting. 

The rest of the paper is organized as follows. In Section \ref{sec:diff:geom}, we provide a brief recap of certain elements of Riemannian geometry for completeness and follow this up in Sections \ref{sec:stoch:dev} and \ref{sec:derivation} with a detailed exposition of the method of stochastic development for SDEs. In Section \ref{sec:applications}, the illustrative applications and the numerical results are provided. A brief discussion on certain future possibilities is also included in the same section. The article is wound up in Section \ref{sec:conclusions} with a few concluding remarks.

\section{Stochastic development of an SDE on Riemannian Manifold }

By way of a ready reference, we give brief reviews of a few basic concepts in  differential geometry and stochastic development in Sections \ref{sec:diff:geom} and \ref{sec:stoch:dev} respectively. In Section \ref{sec:derivation}, we use the notion of frame bundles on Riemannian manifolds to develop an SDE with a non-zero drift.

\subsection{A brief review of concepts from differential geometry}\label{sec:diff:geom}

Differential geometry is the mathematical machinery for performing calculus over an arbitrarily shaped hypersurface in any dimension, say $\mathbb{R}^d$ and can be seen as a useful generalization of standard calculus in the Euclidean setting. The departure from the Euclidean set-up is specifically captured through certain incompatibility tensors, e.g. the curvature tensor in Riemannian geometry. A small neighbourhood around every point in the hypersurface, which is referred to as the manifold, is represented by a local co-ordinate chart, possibly drawn from the embedding Euclidean space. The embedding Euclidean space is of a strictly higher dimension, say $\mathbb{R}^n$ with $n>d$. These local charts overlap smoothly to enable calculations on the manifold as a whole. An important concept in the theory of differential geometry is that of a tangent plane. As the name suggests, it is the unique plane tangent to the manifold at a given point. Formally, a manifold is called Riemannian if the tangent plane at every point $p$ is equipped with an inner product with respect to a given metric $g$ such that, if $X_p$ and $Y_p$ are two vectors on the tangent plane, we have 
\begin{equation}
\langle{X_p,Y_p}\rangle = [g_p]_{ij}x^iy^j
\end{equation}
 where $X_p=x^i e_i \; , Y_p=y^j e_j \; ; \; \{e_i\}_{i=1}^d \text{being the canonical basis vectors in }  \mathbb{R}^d$.
 
In the Euclidean setting, we have $g_{ij} = \delta_{ij}$ where $\delta_{ij}$ represent the Kronecker delta symbols. Loosely speaking, $g$ encapsulates the notion of how distances and angles between two vectors are measured on a tangent plane. It is known that every Riemannian manifold (RM) is associated with a unique Riemannian metric. Now that we have seen that every point on the RM has a tangent plane attached to it and that every tangent plane in turn has a unique metric, one must also figure out a way to smoothly move from one tangent plane to another in a close neighbourhood of the former (parallel transport of vector and tensor fields). This is precisely where the concept of a connection comes in. For a given Riemannian metric $g$, the coordinate representation of the connection is given as 
\begin{equation}
\gamma^k_{ij} = \frac{1}{2}g^{kl}[\partial_ig_{jl} +\partial_jg_{il} -\partial_lg_{ij}]
\end{equation}
In the above equation,  $g^{kl} = g^{-1}_{kl}$ and the symbols $\gamma^k_{ij}$ are also referred to as the Christoffel symbols. It must be noted that $\gamma$ is not a tensor, as it does not transform like one under a smooth change of co-ordinates. The usual concept of derivatives of vectors in $\mathbb{R}^n$ does not apply on the RM, since any two vectors lying in two different tangent planes are objects of different vector spaces, and hence cannot be added or subtracted in the usual way. The equivalent notion of derivative on the RM is known as covariant derivative and it is defined in terms of the connection. The covariant derivative of a vector $Y$ along a vector $X$ in terms of the Christoffel symbols is defined as follows:
\begin{equation}\label{delXY}
\nabla_XY = [XY^k + X^iY^j\gamma^k_{ij}]e_k
\end{equation}
where $X = X^ie_i $, $Y=Y^je_j$, $e_i$ is the unit vector in the $i^{th}$ co-ordinate direction in terms of a local chart. We emphasize that equation (\ref{delXY}) is valid only within the cutlocus; roughly speaking the cutlocus at a point $p$ on the manifold is that neighbourhood (on the manifold) every point in which has a geodesic connecting the point $p$ (see below for the definition of a geodesic on the RM).

Now that we have a way of moving from one point on the manifold to another using the connection, we can define curves. An important example of a curve on the manifold, parametrized by $t$, is that of a geodesic. It is the shortest path joining two given points on the manifold. The equation of a geodesic is as follows:
\begin{equation}
\ddot{x}^k(t) + \dot{x}^i_t\dot{x}^j_t\gamma^k_{ij}(x(t)) = 0
\end{equation}
The Euclidean equivalent of the above equation is just $\ddot{x}^k(t)=0$, solutions to which are straight lines. \\

\subsection{The concept of stochastic development}\label{sec:stoch:dev}
We may combine the basics of stochastic calculus with differential geometry to recast an SDE, originally posed in a $d$ dimensional Euclidean space, on a Riemannian manifold $M$ of the same dimension. A systematic framework for this is provided by stochastic development, which has been used in \cite{HsuElton} to recast a Brownian motion on $M$. We presently use a similar strategy for SDEs that have a non-zero drift. In order to relate the canonical $d$ dimensional Euclidean basis to a basis of the tangent plane $T_xM$ at the point $x\in M$, we need an additional construct of a $d+d^2$ dimensional manifold called the frame bundle $F(M)$. While the $d$-dimensional component of $F(M)$ is the base manifold $M$ itself, the remaining $d^2$-dimensional part corresponds to orthogonal linear transformations applied to vectors on $T_xM$. We now reflect on how the connection $\nabla$ on $M$ manifests itself on the frame bundle $F(M)$. Clearly, a frame at a point $x\in M$ provides a linear isomorphism between the Euclidean space $\mathbb{R}^d$ where the solution of a standard SDE evolves and the $d$-dimensional tangent plane $T_xM$ to $M$ on which the solution needs to be projected. Thus, it is through the frame bundle that we can track these paths on $M$ once we know how it evolves in $\mathbb{R}^d$. 
Let $E_1,...,E_d$ be the co-ordinate basis vectors of the $d$-dimensional Euclidean space. Now considering a frame $q$ at $x$, we note that the vectors $qE_1,...,qE_d$ make up a basis for $T_xM$. 

We denote by $F(M)_x$ the set of all frames at $x$ so that the elements of $F(M)_x$ may be acted upon by $GL(d,\mathbb{R})$, the general linear group. This means that any linear transformation of $F(M)_x$ is also a valid frame at $x$. $F(M)_x$ is also called a fibre at $x$. However, the base manifold $M$ is presently Riemannian so that the torsion tensor is zero, and thus an orthonormal frame remains orthonormal upon parallel transport along $M$. There is thus no loss of generality in restricting the general linear group to the orthogonal group $O(M)$. Roughly speaking, a fibre $\mathcal{F}_x$ at a point $x$ on $M$ is defined as a space attached to that point. We may now define a surjective or onto map $\pi : \mathcal{F(M)}_x \longrightarrow M$. We define the frame bundle as the union of sets of frames at different points on the manifold, i.e.  $F(M) = \bigcup_{x \in M} F(M)_x$. At this stage, we may actually look upon $ F(M)$ itself as a (differentiable) manifold of dimension $d + d^2$. Accordingly, the  projection map $\pi : F(M) \longrightarrow M$ is also smooth.
Now we consider a point $q\in F(M)$ and the associated tangent space $T_qF(M)$ at the same point. It is a vector space of dimension $d+d^2$. We refer to a tangent vector $Y \in T_qF(M)$ as vertical if $Y$ is tangent to the frame $F(M)_{\pi q}$. These vertical tangent vectors form a subspace $V_qF(M)$ of $T_qF(M)$ and it is of dimension $d^2$. Let the base manifold $M$ be equipped with a Riemannian connection $\nabla$. Then a curve $ q_t$ in $F(M)$, which is basically a smoothly varying field of frames, could be projected to a smooth curve $x_t= \pi q_t$ on $M$. We call the frame field $ q_t$ horizontal if the vector field $q_t E$ is parallel along the projected curve $x_t$ on the base manifold $M$ for an arbitrary vector $E \in \mathbb{R}^d$. We recall here that a vector field $V$ along the curve $x_t$ on $M$ is called parallel along $x_t$ if $\nabla_{\dot{x}}V = 0$ for every $t$. This is just an extension of the notion of parallel vectors in the Euclidean setting. The vector $V_{x_t}$ at $x_t$ is the parallel transport of the vector $V_{x_0}$ at $x_0$.

We call a tangent vector $X \in T_qF(M)$ horizontal if it is tangent to the horizontal curve $q_t$. The space of horizontal vectors at $q$ is denoted by $H_qF(M)$; it is a subspace of $T_qF(M)$ and is dimension $d$. We thus have the direct-sum decomposition $$ T_qF(M) = V_qF(M) \oplus H_qF(M)$$
Using the projection $\pi : F(M) \longrightarrow M$, a pushforward operation $\pi_* : H_qF(M) \longrightarrow T_{x}M$ may be defined. Specifically, consider any vector $X \in T_xM$ and a frame $q$ at $x$. The horizontal lift of $X$ is then a unique horizontal vector $X^* \in H_qF(M)$ such that its projection returns the original vector itself, i.e. $\pi_* X^* = X$. Now consider any Euclidean vector $E \in \mathbb{R}^d$. The vector $H_E(q)$ at the point $q$ in $F(M)$ is defined by the horizontal lift of the vector $qE$ on $M$, i.e. $H_E(q) = (qE)^* $. Hence, $(qE)^* $ may be interpreted as a horizontal vector field on $F(M)$. Corresponding to the unit (orthonormal) coordinate vectors $E_1,...,E_d$ in $\mathbb{R}^d$, we note that $H_i := H_{E_i}, \;\; i=1,...,d$, are the associated horizontal vector fields of the frame bundle that span the horizontal subspace $H_qF(M)$ at each $q \in F(M)$.

We may adopt any valid local chart $x=\{x^i\}$ in a  neighbourhood $O \subset M$. Using the inverse of the projection map, this local chart on the base manifold $M$ induces a local chart $\tilde{O} = \pi^{-1}(O)$ in $F(M)$. Thus, let $X_i = \frac{\partial}{\partial x^i} , \;\; 1 \leq i \leq d $, be the coordinate basis vectors.  For a frame $q \in \tilde{O} $, we have $qE_i = Q^j_i X_j$ for some matrix $Q = (Q^i_j)$. Accordingly, we get $(x,q) \in \mathbb{R}^{d+d^2}$ as the local chart for $\tilde{O}$. Then, the vertical subspace $V_qF(M)$ is spanned by $X_{kj} = \frac{\partial}{\partial Q^k_j}, \;\; 1 \leq j,k \leq d$. Also, the vector fields $\{X_i,X_{ij}, 1 \leq i,j \leq d\}$ span $T_qF(M)$, $q \in \tilde{O}$. An expression  for the horizontal vector field $H_i$ in terms of the local coordinates is given as follows. 
\begin{equation} \label{horz-vector}
H_i(q) = Q^j_iX_j - Q^j_i Q^l_m \gamma^k_{jl}(x) X_{km}
\end{equation}
For the sake of brevity, we skip the proof here and refer to (\cite{HsuElton}).

From the definition of $q_t$, which is the horizontal lift of a smooth curve $x_t$ on $M$, we have $q_t^{-1}\dot{x}_t \in \mathbb{R}^d $ since $\dot{x}_t \in T_{x_t}M$. We define the anti-development of $\{x_t\}$ on $M$ as a curve $ u_t$ in $\mathbb{R}^d$ such that the following equation is satisfied. 
$$u_t = \int_0^t q_s^{-1} \dot{x}_s ds  .$$
In other words, $q_t \dot{u}_t = \dot{x}_t$ and by the definition of horizontal vector fields, we have $H_{\dot{u}_t}(q_t) = (q_t \dot{u}_t)^* = (\dot{x}_t)^* = \dot{q_t} $, i.e. the anti-development $u_t$ and the horizontal lift $ q_t$ of a curve $x_t$ on $M$ are simply related by an ordinary differential equation (ODE). In view of our work in the next subsection, it is expedient to rewrite the last equation as 
\begin{equation}\label{eqn:sdode}
    \dot{q_t}=H_i(q_t) \dot{u_t}
\end{equation}
If we start from an Euclidean curve $u_t$ in $\mathbb{R}^d$ and a frame $q_0$ at the point $x_0$ on $M$, the unique solution of the above ODE is given by a horizontal curve $q_t$ in $F(M)$. We refer to this horizontal curve as the development of $u_t$ in the frame manifold $F(M)$. Its projection on $M$ given by $\pi q_t$ is called the development of $u_t$ in $M$.

\subsection{Local coordinate expression of a developed SDE on RM}\label{sec:derivation}
We extend equation \ref{eqn:sdode} to the stochastic case and write it in the Stratonovich sense as:  
\begin{eqnarray}
dq_t = H_iq(t) \circ dW^i_t 
\end{eqnarray}
where the Ito SDE for the Euclidean stochastic process $W_t$ has the following form: 
\begin{equation}
dW^i_t = \alpha^i(W_t)dt + \beta^i_j(W_t)dB^j_t
\end{equation}
From \cite{HsuElton} (see proposition 2.1.3), the horizontal vector fields are locally given by the equation below.
\begin{equation}
H_i(q) = Q^i_jX_j - Q^i_jQ^l_m \gamma^k_{jl} X_{km} 
\end{equation}
where 
\begin{equation}
X_i = \frac {\partial}{\partial x^i} \;\;\; X_{km}= \frac {\partial}{\partial Q^k_m}
\end{equation}
Hence, written in the Stratonovich sense, the equation for $q_t = \{x^i_t,Q^i_j(t)\}$  is
\begin{eqnarray}
dx^i_t &=& Q^i_j(t) \circ dW^j_t \label{dx:eqn} \\[15pt]
dQ^i_j(t) &=& -  \gamma^i_{kl}(x_t)Q^l_j(t)Q^k_m(t) \circ dW^m_t \label{de:eqn}
\end{eqnarray}
From equation (\ref{dx:eqn}) and in the Ito sense, we have
\begin{eqnarray}
dx^i_t &=& Q^i_j(t) dW^j_t + \frac{1}{2}d \langle{Q^i_j(t) ,dW^j_t }\rangle \label{dX:eqn} \\[15pt]
&=& Q^i_j(t)\alpha^j(W_t)dt + Q^i_j(t) \beta^j_m(W_t)dB^m_t + \frac{1}{2}d \langle{Q^i_j(t) ,dW^j_t }\rangle    \nonumber
\end{eqnarray}
Let $dM^i_t = Q^i_j(t) \beta^j_m(W_t)dB^m_t  \;$ be the martingale part. Then we have
\begin{eqnarray}
d\langle{M^i_t,M^j_t }\rangle &=& \langle{Q^i_m(t) \beta^m_r(W_t)dB^r_t  ,Q^j_n(t) \beta^n_s(W_t)dB^s_t  }\rangle \\[15pt]
&=& Q^i_m(t) \beta^m_r(W_t)Q^j_n(t) \beta^n_r(W_t) dt
\end{eqnarray}
However, we have $qE_l = Q^i_lX_i$ and 
$\delta_{lm} = \langle{qE_l,qE_m}\rangle =\langle{Q^p_lX_p , Q^q_mX_q}\rangle  = Q^p_lQ^q_m\langle{X_p , X_q}\rangle =g_{pq}Q^p_lQ^q_m $.
Thus, $QgQ^T=I$ or $Q^TQ=g^{-1}$. Accordingly, we may write
\begin{equation}
d\langle{M^i_t,M^j_t }\rangle = [\beta^T g^{-1} \beta]_{ij}dt
\end{equation}
Now, let $\sigma = Q\beta$. Then 
\begin{eqnarray}
dM^i_t&=&[Q\beta]_{im}dB^m_t \\[15pt]
&=&\sigma_{im}dB^m_t 
\end{eqnarray}
From equation (\ref{de:eqn}), we have
\begin{equation}
dQ^i_j(t) = -\gamma^i_{kl}(x_t)Q^l_j(t)Q^k_m(t)dW^m_t + \frac{1}{2}d\langle{-\gamma^i_{kl}(x_t)Q^l_j(t)Q^k_m(t),dW^m_t}\rangle
\end{equation}
Thus, the last term on the RHS of equation (\ref{dX:eqn}) becomes
\begin{eqnarray}
d\langle{Q^i_j , dW^j_t}\rangle &=&\langle{dQ^i_j , dW^j_t}\rangle \\[15pt]
&=&\langle{ -\gamma^i_{kl}(x_t)Q^l_j(t)Q^k_m(t)dW^m_t, dW^j_t}\rangle \nonumber \\[15pt]
&=& \langle{ -\gamma^i_{kl}(x_t)Q^l_j(t)Q^k_m(t)[\alpha^m(W_t)dt + \beta^m_p(W_t)dB^p_t], [\alpha^j(W_t)dt + \beta^j_q(W_t)dB^q_t]}\rangle \nonumber \\[15pt]
&=& -\gamma^i_{kl}(x_t)Q^l_j(t)Q^k_m(t) \beta^m_p \beta^j_q \langle{dB^p_t,dB^q_t}\rangle \nonumber \\[15pt]
&=& -\gamma^i_{kl}(x_t)Q^l_j(t)Q^k_m(t) \beta^m_p \beta^j_p dt \nonumber  \\[15pt]
&=& -\gamma^i_{kl}(x_t)[Q\beta]_{lp}[Q\beta]_{kp}dt \nonumber \\[15pt]
&=& -\gamma^i_{kl}(x_t) [(Q\beta)(Q\beta)^T]_{kl}dt \nonumber \\[15pt]
&=& -\gamma^i_{kl}(x_t) [\sigma \sigma^T]_{kl}dt \nonumber 
\end{eqnarray}
Substituting in equation (\ref{dX:eqn}), we finally get the developed SDE.
\begin{equation}\label{GALE}
dx^i_t =  [\sqrt{g^{-1}(x_t)}]_{ij}\alpha^j(W_t)dt + \sigma_{im}(W_t)dB^m_t -\frac{1}{2}  [\sigma \sigma^T]_{kl} \gamma^i_{kl}(x_t) dt
\end{equation}
where $\sigma = \sqrt{g^{-1}} \beta$.

\section{Applications and illustrations}\label{sec:applications}
We now demonstrate how the developed flows of stochastic dynamical systems could be meaningfully exploited to arrive at significantly improved numerical approaches for a broad range of applications. These include simulations of a Brownian particle trapped in a potential well, a numerical integration scheme that can preserve the mean-energy drift for a stochastic Hamiltonian flow under additive noise and a stochastic search scheme for non-convex optimization. In all these illustrations, the developed SDEs on the RM are integrated by a most basic version of the explicit Euler-Maruyama (EM) scheme with a strong error order $\mathcal{O}(\sqrt{\Delta t})$, where $\Delta t$ is the integration step size, presently assumed to be uniform. To showcase the improvement, a solution through the geometric approach is always compared with that of the standard SDE in the Euclidean setting -- both integrated via the explicit EM method.    

\subsection{Brownian motion in a potential well}

Brownian motion in a potential well is widely studied to understand  myriad phenomena at the molecular level, e.g. to model the deterministic and stochastic forces at play. This is also the underlying principle for optical and acoustic tweezers. The trapping of Brownian particles via optical/acoustic tweezers has proved pivotal in the experimental understanding of numerous phenomena in science and engineering, and this is an important development that cannot be realized with unrestricted BM. First introduced in \cite{Ashkin1970}, the simplest application of an optical tweezer is to laser-trap a single Brownian particle, viz. a dielectric object of the size of a nanometer to a micrometer \cite{ChuAshkin1986}. This is typically done for molecular motion or force measurements or for non-invasive manipulations of a single cell. Considerable work has been reported on optical tweezers; see \cite{ashkin1997optical} for a review. However, a similar non-contact immobilization of cells or particles in microfluidic systems is also possible with acoustic traps \cite{lee2009single}, where ultrasound standing waves are used for trapping purposes. Acoustic traps are known to be safer and hence more suitable for biological applications, especially as optical traps may kill some organisms to be studied due to excessive heating from lasers.

\begin{figure}
     \centering
     \begin{subfigure}[b]{0.45\textwidth}
         \centering
         \includegraphics[width=\textwidth]{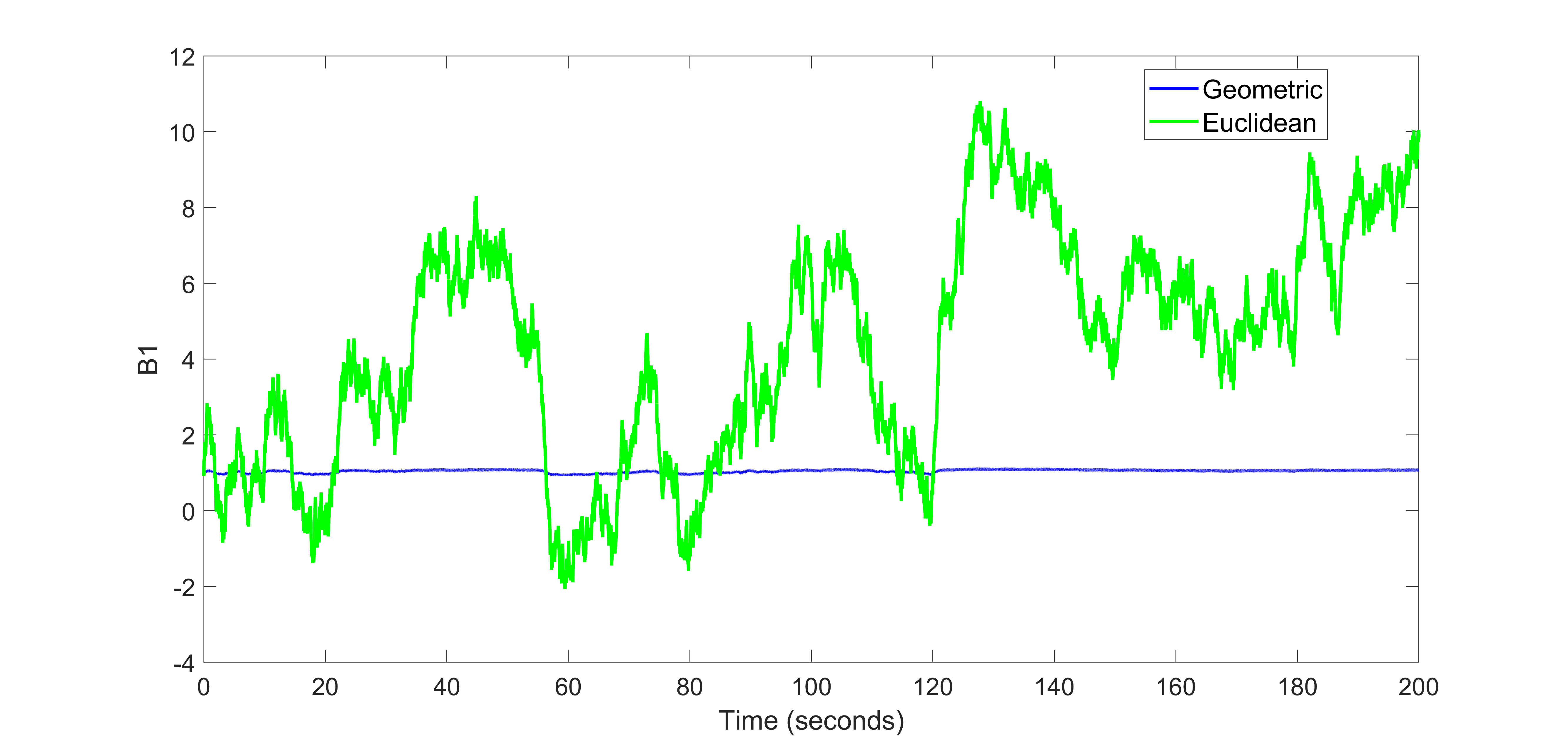}
         \caption{}
         \label{fig:B1}
     \end{subfigure}
     \hfill
     \begin{subfigure}[b]{0.45\textwidth}
         \centering
         \includegraphics[width=\textwidth]{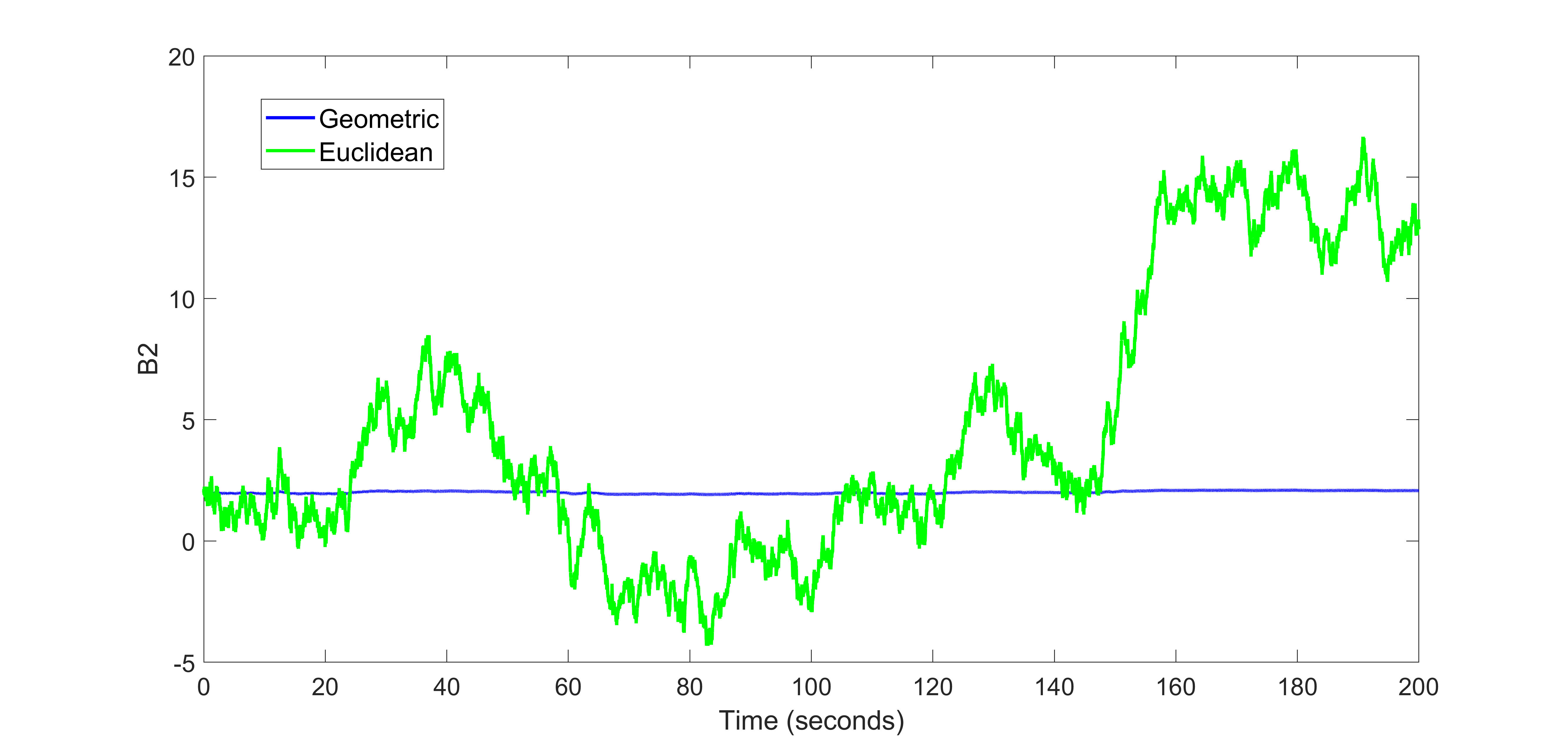}
         \caption{}
         \label{fig:B2}
     \end{subfigure}
        \hfill
     \centering
     \begin{subfigure}[b]{0.45\textwidth}
         \centering
         \includegraphics[width=\textwidth]{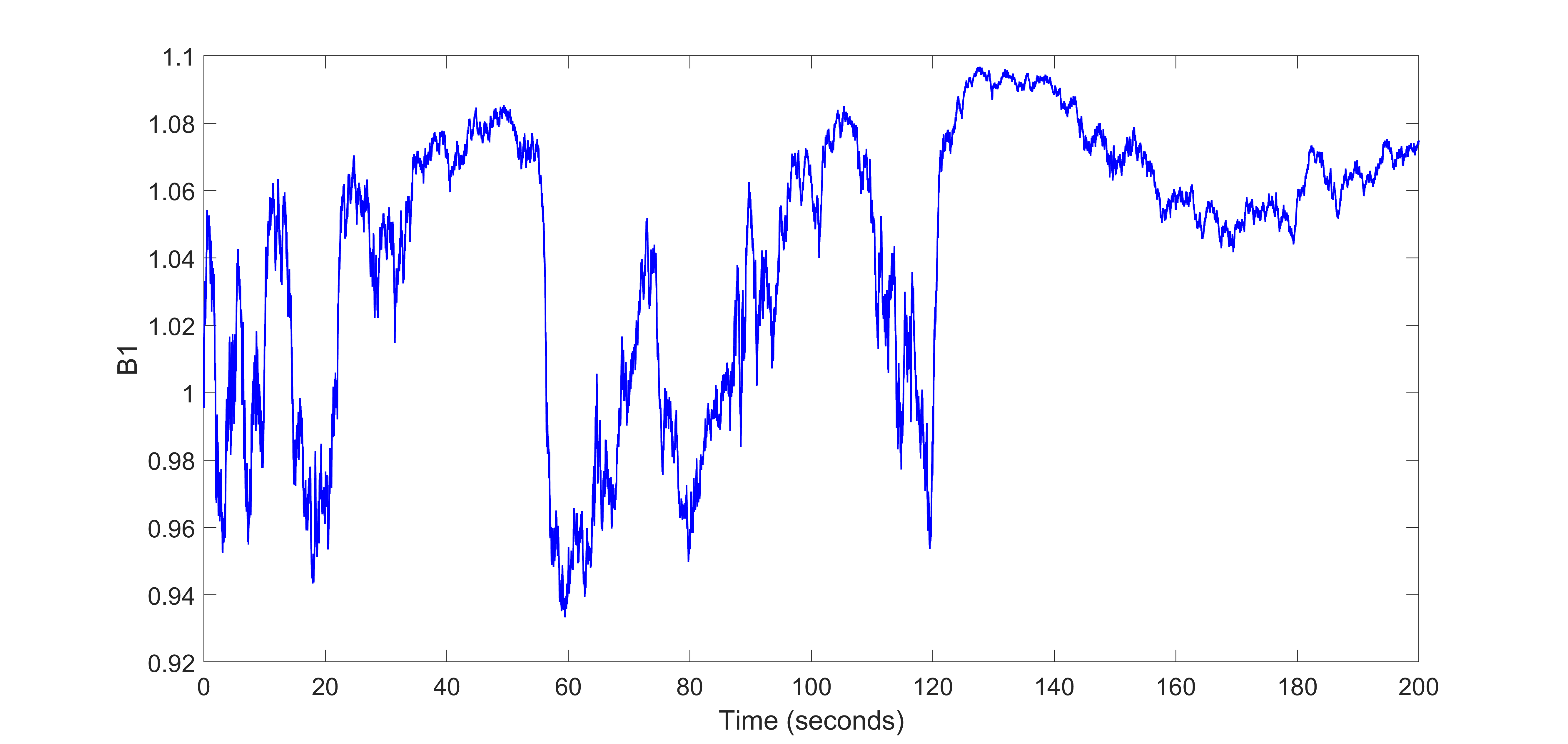}
         \caption{}
         \label{fig:B1}
     \end{subfigure}
     \hfill
     \begin{subfigure}[b]{0.45\textwidth}
         \centering
         \includegraphics[width=\textwidth]{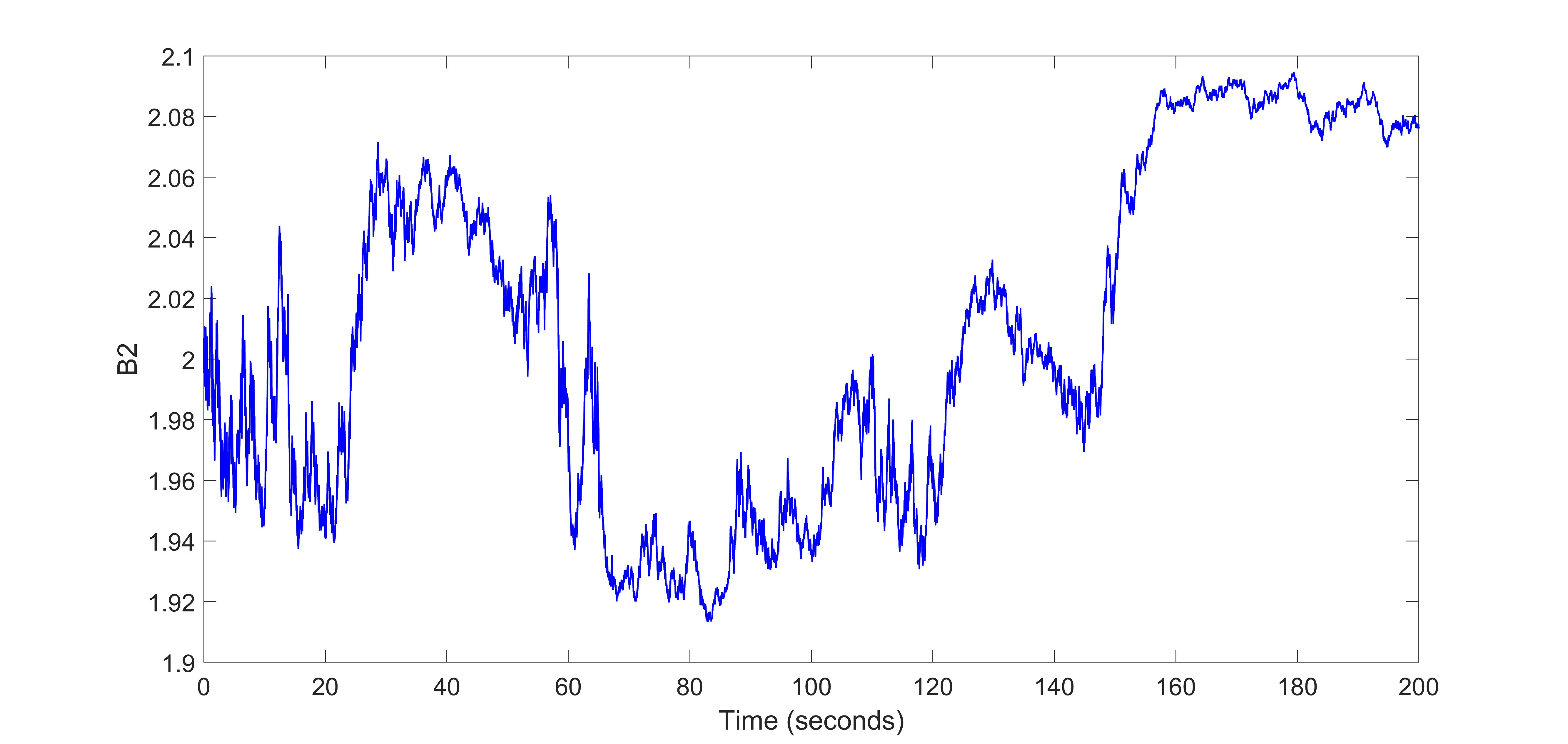}
         \caption{}
         \label{fig:B2}
     \end{subfigure}
        \hfill
     \centering
     \begin{subfigure}[b]{0.45\textwidth}
         \centering
         \includegraphics[width=\textwidth]{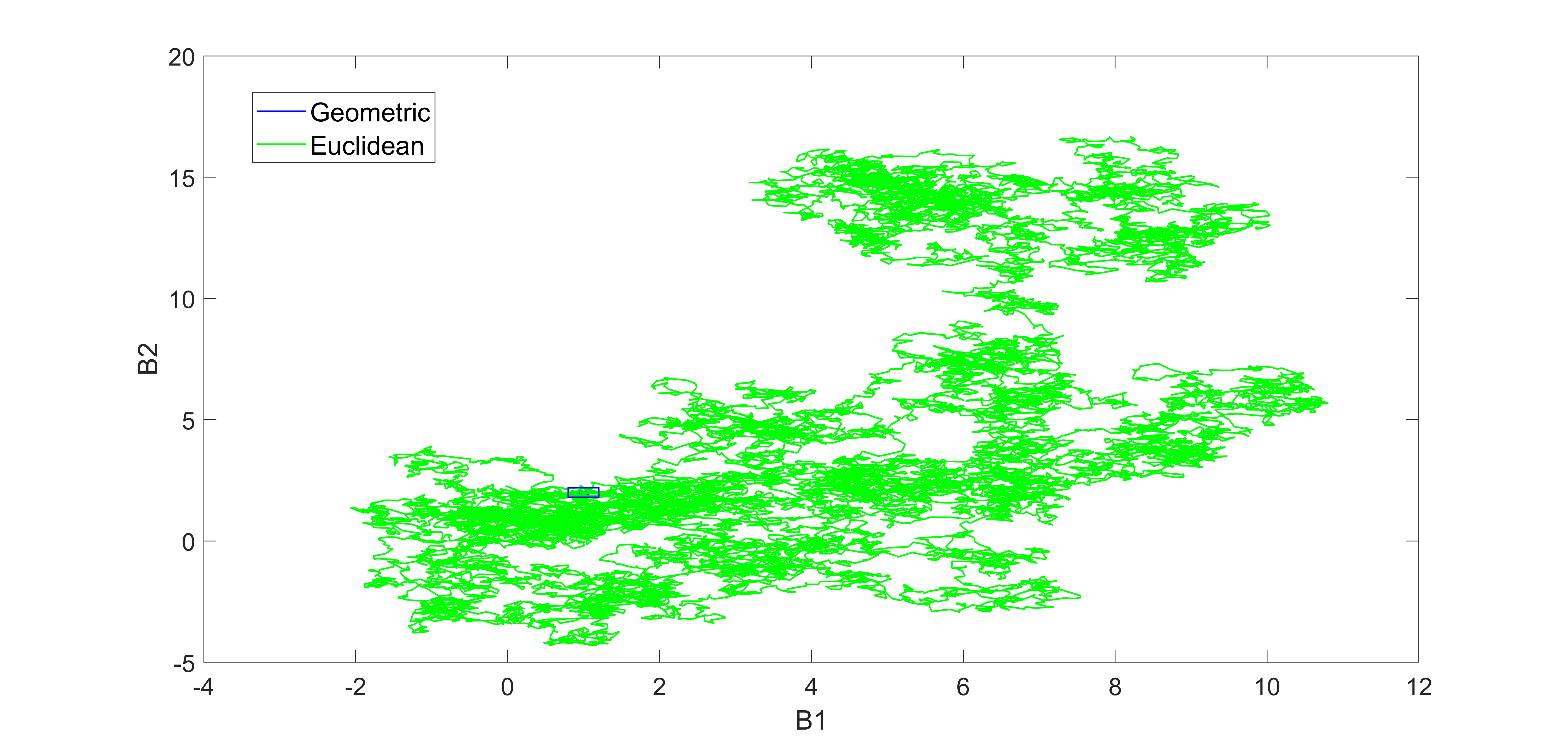}
         \caption{}
         \label{fig:B1}
     \end{subfigure}
     \hfill
     \begin{subfigure}[b]{0.45\textwidth}
         \centering
         \includegraphics[width=\textwidth]{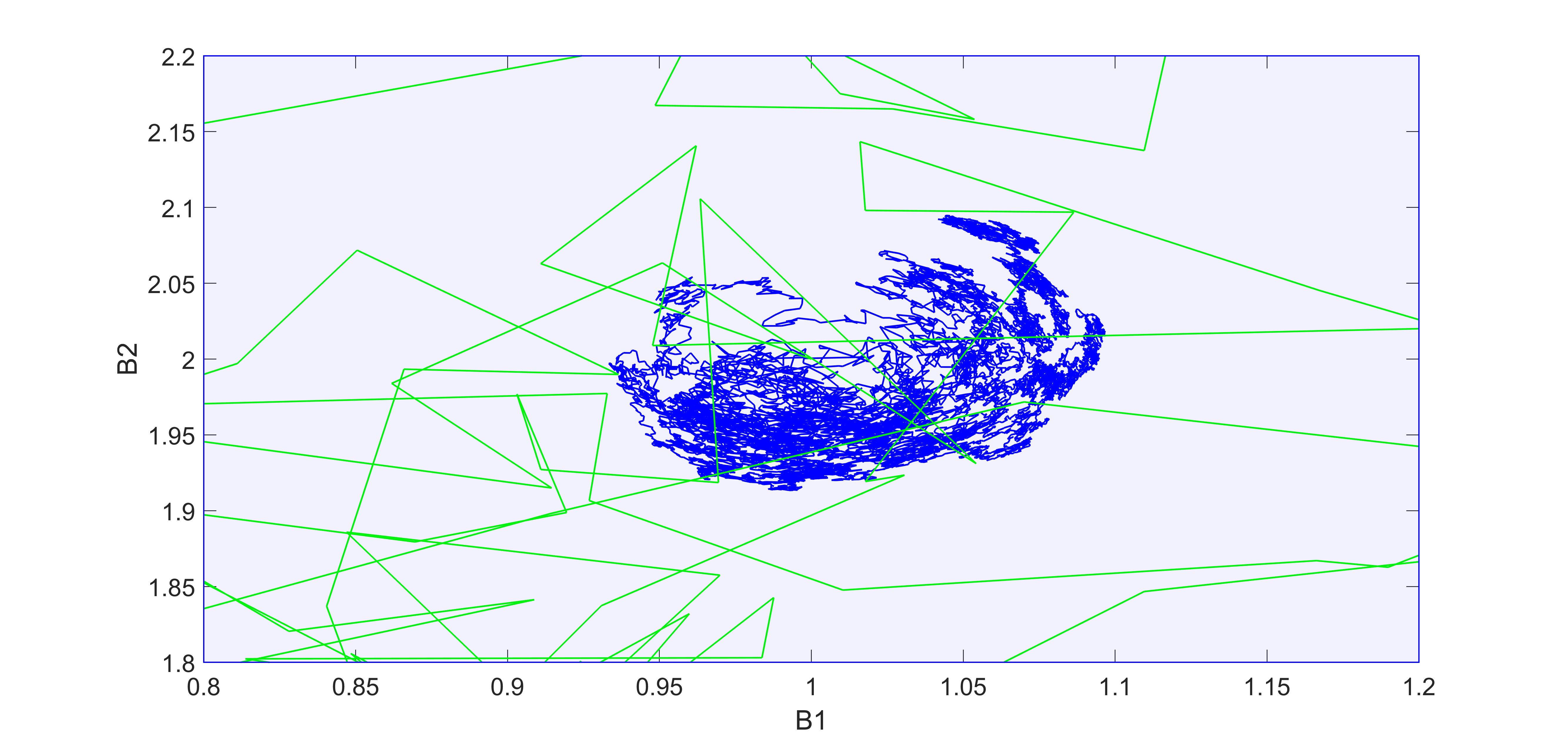}
         \caption{}
         \label{fig:B2}
     \end{subfigure}
        \caption{ Comparison between the geometric and Euclidean methods for the time history of Brownian motion components $B_1$ and $B_2$ are shown in (a) and (b) respectively. Time history of only geometric Brownian motion is shown in (c) and (d) for clarity. Comparison of Brownian motion path in space is shown in (e), while (f) shows the inset from (e) for the geometric method. Problem parameters: $\Delta t = 0.01$, $d_1=d_2=400$, the center of potential well assumed as $\beta_1=1,\beta_2=2$}
        \label{fig:BM:potential:well}
\end{figure}

Modelling of these tweezers requires that the equation of motion of a Brownian particle be trapped in a potential well. One way of simulating such motion is to apply Doob's h-transform \cite{RoyDebasish}, where an appropriate drift term to trap the Brownian particle could, in principle, be found based on a change of measures. Implementing this within a numerical approach is however quite formidable and requires an accurate inversion of the heat kernel. We presently simulate such a Brownian motion via equation \ref{GALE} by requiring that the original drift field of the Euclidean SDE be zero. This is also the well-known equation for Brownian motion on an RM available in the literature \cite{HsuElton}. The Riemannian metric and the connection for this are arrived at from the expression of the potential well. Equations for the potential well $E$ and the Riemannian metric $g$ are given below. The associated Levi-Civita connection may be derived from the expression for $g$; see Appendix A.

\begin{equation}
    E(x) = \exp{(x-\lambda)^T [\alpha] (x-\lambda)}
\end{equation}
Assuming that $\alpha$ is a diagonal matrix with entries $d_1, d_2...d_n$ where $n$ is the dimension of $x$, we have:
\begin{equation}
    g_{ij}  = \frac{1}{2} \frac{\partial^2 E(x)}{\partial x_i \partial x_j}= 2 d_{(i)} d_{(j)} (x-\lambda)_{(i)} (x-\lambda)_{(j)} \exp(d_p (x-\lambda)_p^2) + d_{(j)} \delta_{ij} \exp(d_p (x-\lambda)_p^2) 
\end{equation}
Note that the indices in brackets imply no sum. The developed SDE corresponding to the Euclidean SDE $dX_t = dB_t$ is as follows:

\begin{equation}\label{eqn:BMwell}
    dX^i_t = dB^i_t - \frac{1}{2}g^{-1}_{kl} \gamma^i_{kl} dt
\end{equation}
The results from our numerical simulations are shown in Figure \ref{fig:BM:potential:well}, where they are compared with the standard Euclidean Brownian motion. Exploiting the metric as well as the connection, the developed SDE \ref{eqn:BMwell} restricts the Brownian dynamics close to the potential well and this feature is clearly brought forth in the figure.  

Since we are dealing with solutions of developed SDEs, the metric defined through certain energy criteria could lose positive definiteness owing to the random fluctuations. One way to address this issue could be based on an additive regularization as outlined below. Suppose that we were to start from an ensemble of random initial conditions, given by the vector valued random variable $X_0$ with density $p(\xi)$. If we take $-\log p(\xi)$ as an energy-like potential, then its Hessian given by $[g_0]_{ij}=-\frac{\partial^2 \log p(\xi)}{\partial \xi_i \partial \xi_j}$ could be taken as the additive regularizer to our original metric $g$. Specifically, if $p(\xi)$ is multivariate Gaussian with mean $\mu_0$ and covariance $\Sigma_0$, then we have $[g_0]_{ij}={\Sigma_0}^{-1}$. A particularly expedient choice, which we frequently use in the examples to follow, is the uncorrelated case given by ${\Sigma_0}^{-1}=\Upsilon I$, where $\Upsilon$ is a positive real and $I$ the identity matrix. 

\subsection{Energy-drift preserving numerical integration of SDEs}
The SDEs arising in scientific and engineering applications typically have a drift field that often contains important information on the underlying energetics. This is particularly so for Hamiltonian systems that are extensively used in myriad applications. When such systems are strictly deterministic and not explicitly time dependent, the energy (i.e. the Hamiltonian) appears as a first integral of motion which is a constant in time. These systems are also symplectic, where their motion preserves the phase space area given by the symplectic two-form. In the absence of dissipation, e.g. damping, errors in the numerical integration of such systems could quickly increase. There are several schemes that are either symplectic or energy conserving during time recursion \cite{yoshida1990construction, channell1990symplectic, ishimori2008high}. As an extension of this line of work to the stochastic case, numerical schemes to preserve the symplectic structure for Hamiltonian systems under additive noise have been reported in \cite{milstein2002symplectic}. A few studies on the preservation of certain integral invariants, e.g. energy, in such systems are also available; see \cite{cohen2014energy}. This last class of extensions has typically considered Hamiltonian dynamics under appropriate multiplicative noises and the methods have been proposed in the Stratonovich sense wherein one may exploit many features of similar schemes used in the deterministic setting. It may be shown using Ito's formula that, under an additive noise with a constant intensity, the mean energy of a Hamiltonian system grows linearly in time \cite{chen2020drift}. Our purpose here is to show that an exploitation of stochastic development could be used to expediently impose a known constraint on the mean energy growth within the numerical integration scheme.

Consider, for instance, a Duffing oscillator which is undamped and unforced except for an additive noise. The equation of motion here is:
\begin{equation}
    \ddot{x} + kx + \alpha x^3 = \sigma \dot{B_t}
\end{equation}
where $k$ and $\alpha$ are the mechanical stiffness parameters associated respectively with the linear and cubic terms in $x$, and $\sigma$ is the diffusive noise coefficient, presently assumed constant. Note that $\dot{B_t}$ is not a valid function as $B_t$ is almost nowhere differentiable. It is more appropriately rewritten in the incremental state space form with $X_{1,t}:=x(t)$ and $X_{2,t}: = \dot{x}(t)$. We thus have the following SDE:
\begin{eqnarray}
 dX_{1,t} &=& X_{2,t} dt \\[15pt]
 dX_{2,t} &=& (-k X_{1,t}  - \alpha X_{1,t}^3)dt  + \sigma dB_t \nonumber
 \end{eqnarray}
For $\sigma =0$, the energy of the oscillator above is a constant of motion given by $H(x_{1,t}, x_{2,t})=\frac{1}{2}{x_{2,t}}^2+\frac{k}{2}{x_{1,t}}^2+\frac{\alpha}{4}{x_{1,t}}^4=H_0$, where $H_0$ is the initial energy that depends on the initial conditions alone. However, for $\sigma \neq 0$, the mean of the energy increases linearly in time and is given by $\mathbb{E}[H(x_{1,t}, x_{2,t})]=\mathbb{E}[H(x_{1,0}, x_{2,0})]+\frac{1}{2}{\sigma}^2t$. Now, the numerical integration of an SDE requires integrating $dB_t$ over a finite step size and the variance of this term grows linearly in $\Delta t$. This is quite in contrast with the discretized drift terms whose variance increases quadratically in $\Delta t$. It is also known that, unlike deterministic ODEs for which many higher order integration schemes are available, such schemes are scarce for SDEs. The difficulty arises in dealing with the multiple stochastic integrals in the Ito-Taylor expansion that forms the basis of any numerical integration scheme. For instance, the strong error order in the explicit EM scheme, which we use in this work, is just $\mathcal{O}(\sqrt\Delta t)$. Indeed, the same error order will formally continue to hold even when we use it to solve the SDE developed on the RM. Even so, as we shall soon see, the geometric route can drastically improve the qualitative nature of the numerical solution. 

Similar to the previous example on Brownian motion in a potential well, we make use of the following energy-like term to constrain the flow around the linearly drifting mean energy.
\begin{equation}
E_t =   \exp{\beta(\frac{x_{2,t}^2}{2}+ \frac{kx_{1,t}^2}{2} + \frac{\alpha x_{1,t}^4}{4} - Z_t)^2} -1
\end{equation}
where
\begin{equation}\label{eqn:theoretical:H}
Z_t = \mathbb{E}[H(x_{1,t}, x_{2,t})]=\mathbb{E}[H(x_{1,0}, x_{2,0})] + \frac{1}{2} \sigma^2 t = H_0 + \frac{1}{2} \sigma^2 t
\end{equation}
Derivations of the metric as well as the connection are on the same lines as in the last illustration (see Appendix B for details). The developed SDE takes the form:
\begin{equation}
    dX^i_t = \sqrt{g^{-1}}_{ij}f^j(X_t)dt + [\sqrt{g^{-1}} \Sigma]_{ij} dB^j_t - \frac{1}{2} [\sqrt{g^{-1}}\Sigma \Sigma^T \sqrt{g^{-1}}^T]_{jk} \gamma^i_{jk} dt
\end{equation}
where $$ f(X_t) = \begin{bmatrix}
X_{2,t} \\
-k X_{1,t} - \alpha X_{1,t}^3
\end{bmatrix}$$
 $\Sigma = \begin{bmatrix} 
0 & 0 \\
0 & \sigma
\end{bmatrix}$
and $\sqrt{g^{-1}}$ denotes the matrix square root of $g^{-1}$. 

\begin{figure}
     \centering
     \begin{subfigure}[b]{0.45\textwidth}
         \centering
         \includegraphics[width=\textwidth]{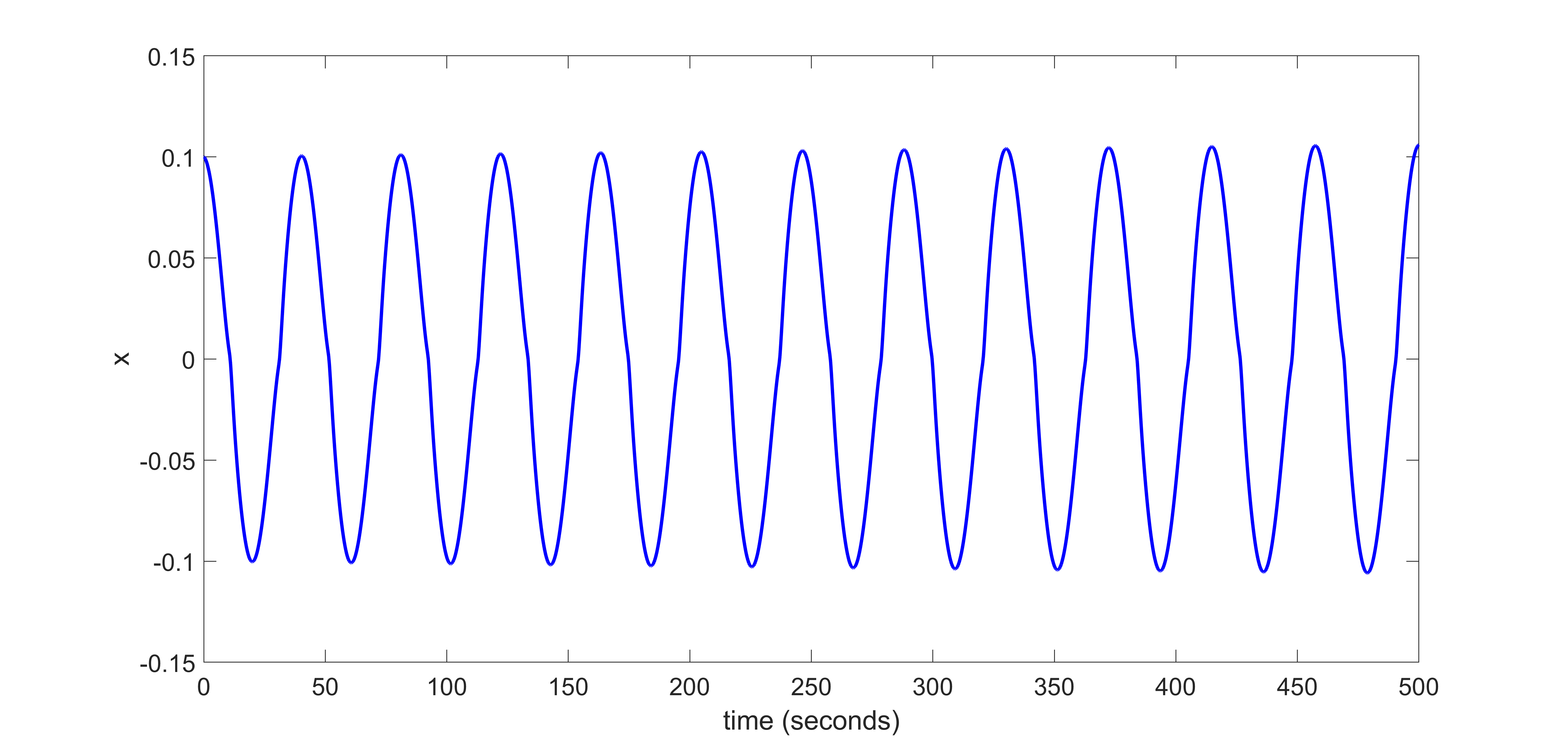}
         \caption{}
         \label{fig:driftp:x}
     \end{subfigure}
     \hfill
     \begin{subfigure}[b]{0.45\textwidth}
         \centering
         \includegraphics[width=\textwidth]{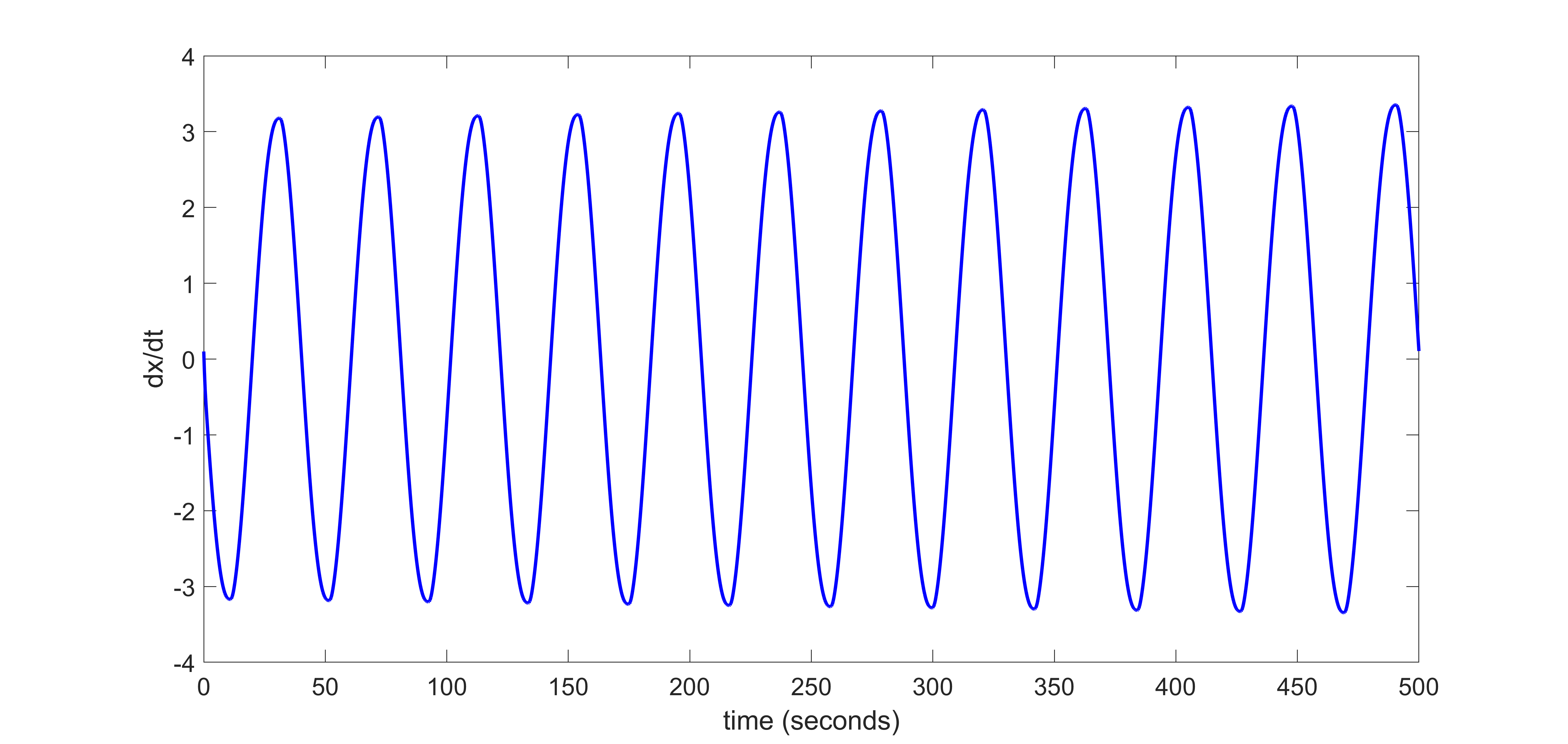}
         \caption{}
         \label{fig:driftp:vel}
     \end{subfigure}
    \hfill
     \begin{subfigure}[b]{0.45\textwidth}
         \centering
         \includegraphics[width=\textwidth]{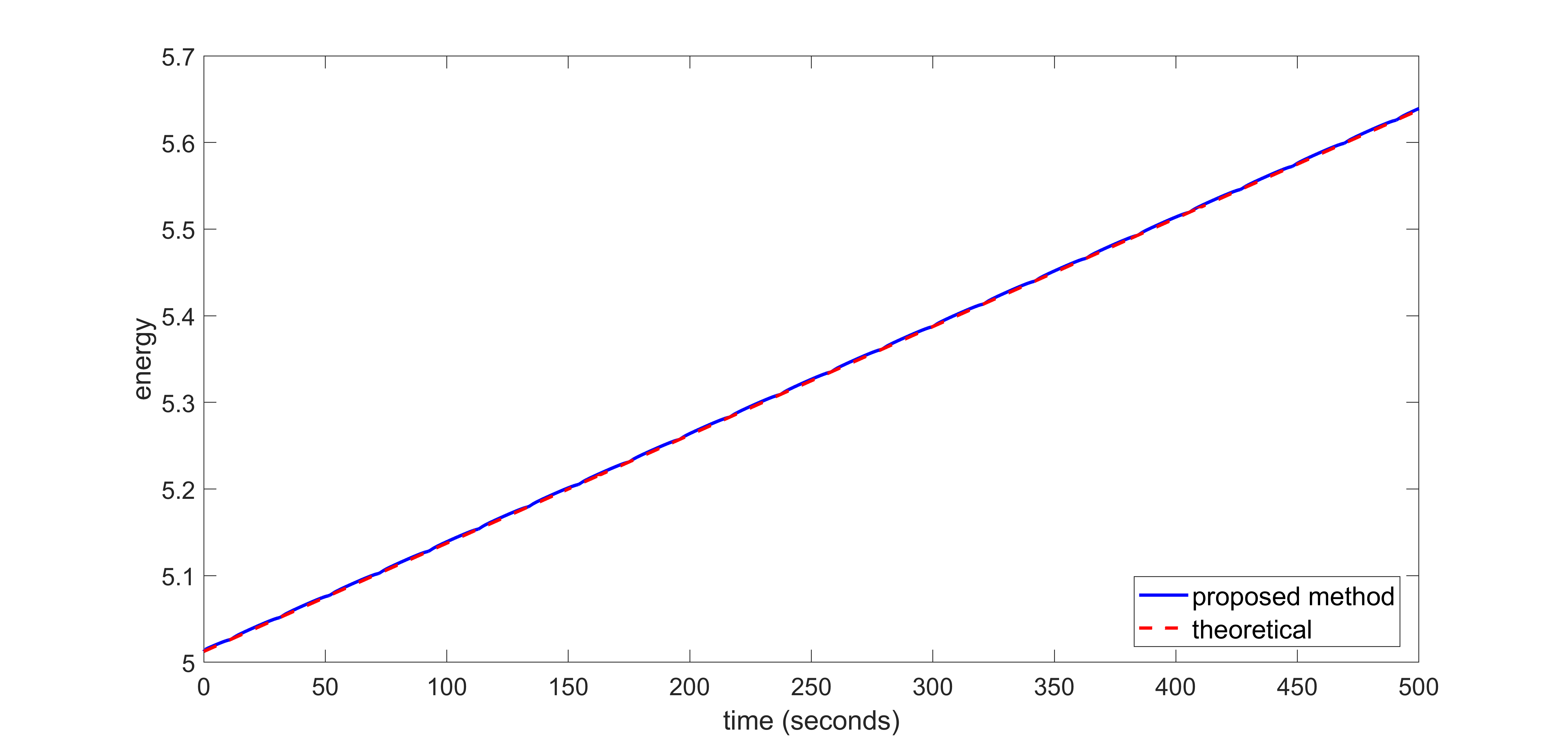}
         \caption{}
         \label{fig:driftp:energy}
     \end{subfigure}
     \hfill
     \begin{subfigure}[b]{0.45\textwidth}
         \centering
         \includegraphics[width=\textwidth]{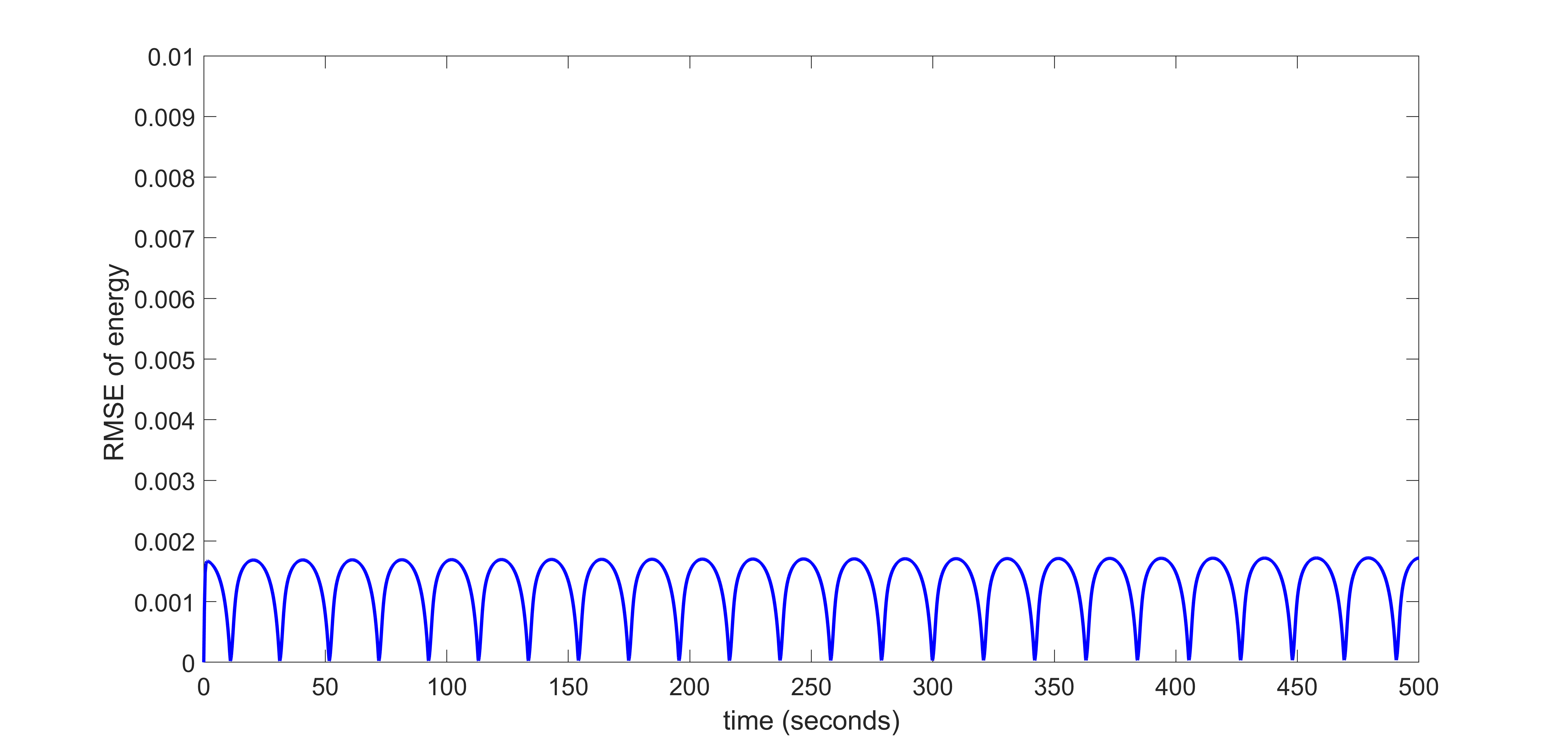}
         \caption{}
         \label{fig:driftp:error}
     \end{subfigure}
     \hfill
     \begin{subfigure}[b]{0.45\textwidth}
         \centering
         \includegraphics[width=\textwidth]{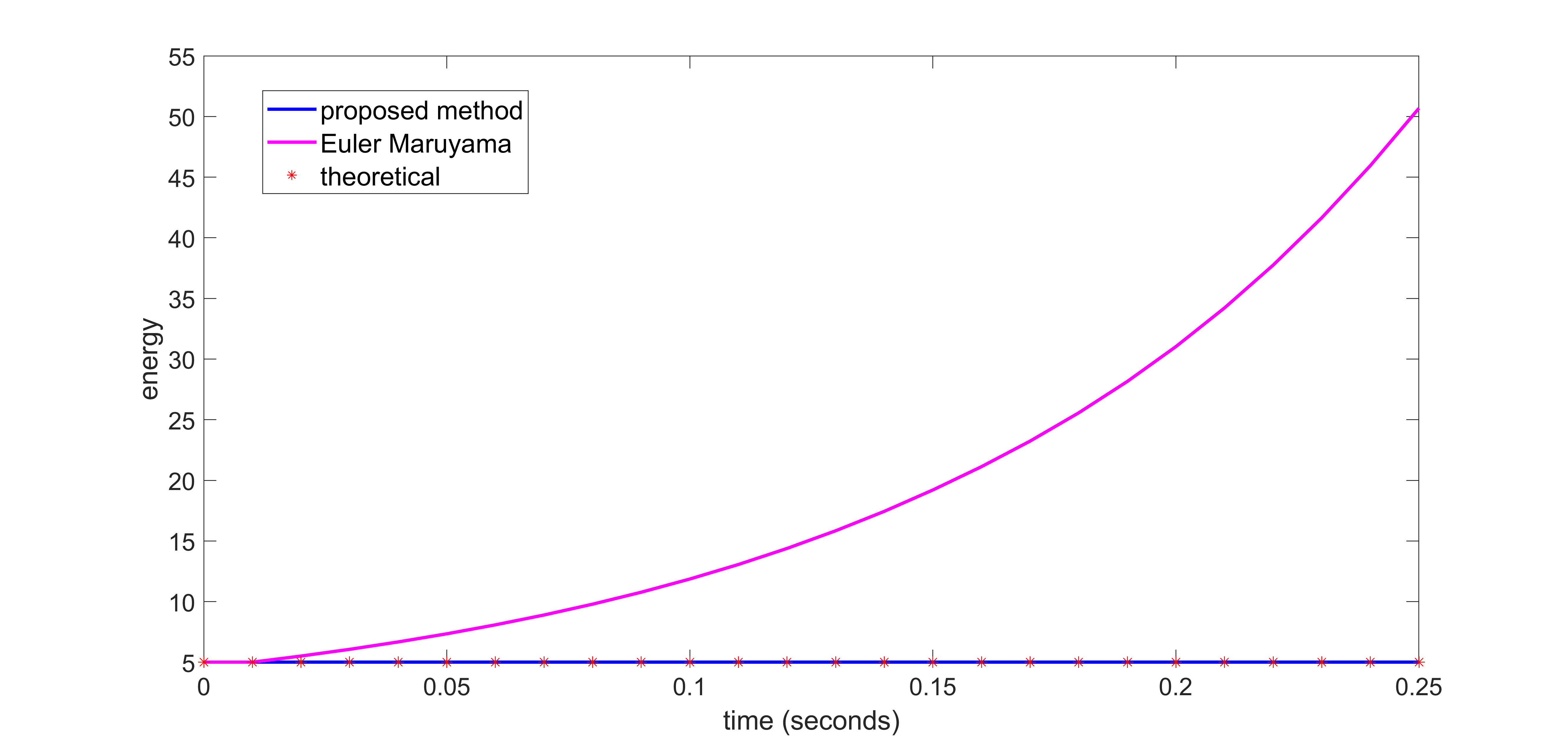}
         \caption{}
         \label{fig:driftp:compare:energy}
     \end{subfigure}
        \caption{Drift preserving integration based on stochastic development. Subfigures (a) and (b) show the displacement and velocity plots respectively, while (c) and (d) show the energy and the error in energy with respect to theoretical. Figure (e) gives the comparison of energy values between the EM and the proposed methods. The parameters are : ensemble size $N=50,\Delta t=0.01, k=1000, \alpha = 300, \sigma = 0.05$, the regularization parameter $\Upsilon = 10^4$,  the initial conditions are $X_{1,0}=0.1, X_{2,0}=0.1$}
        \label{fig:driftp}
\end{figure}

We continue to use an explicit EM scheme to numerically integrate the original and developed SDEs; the results are reported in Figure \ref{fig:driftp}. The displacement and velocity curves as obtained by integrating the developed SDEs are shown respectively in Figures 2(a) and 2(b); the results via the standard EM integration scheme blow off quickly and hence not shown. Figure 2(c) shows a comparison of the theoretical energy (see equation \ref{eqn:theoretical:H}) versus the one based on the proposed method; the RMSE plot for the same is reported in Figure 2(d). Figure 2(e) depicts the energy plots via the proposed and EM methods over a relatively shorter initial time window. The vastly superior performance of the geometric method is self evident.

\subsection{Non-convex optimization}
 
 In this section, we consider the application of stochastic development to an optimization problem that involves a non-convex objective function. The aim of our optimization scheme is then to minimize this function. Within a stochastic search framework, we specifically do this by developing the overdamped Langevin SDE whose evolution is additionally guided by a simulated annealing procedure. In this context, note that a strictly positive, smooth, scalar-valued and non-convex objective function $f(x)$ could be looked upon, at least locally, as an energy-like functional in the space of the design variables. Now consider, for example, the minimization of the Ackley function which constitutes one of the benchmark problems \cite{tang2007benchmark}, often used to test the performance of an optimization scheme. Treating $f(x)$ as the energy, we may readily determine $g$ and $\gamma$; see Appendix C for details. During a stochastic search involving a non-convex function, $g$ may sometimes become negative-definite, particularly during the initial stages. As noted before, we use an additive regularizer of the type $\Upsilon I$ in order to ensure positive-definiteness of $g$. We then use the developed SDE for the overdamped Langevin dynamics with simulated annealing to carry out the evolutionary search for the global minimum of $f(x)$. The results so obtained are also contrasted with those via the overdamped Langevin dynamics with simulated annealing, but without stochastic development. One may note that the simulated annealing step expedites a more exhaustive search of the design space during the initial stages.   

\begin{figure}
     \centering
     \begin{subfigure}[b]{0.45\textwidth}
         \centering
         \includegraphics[width=\textwidth]{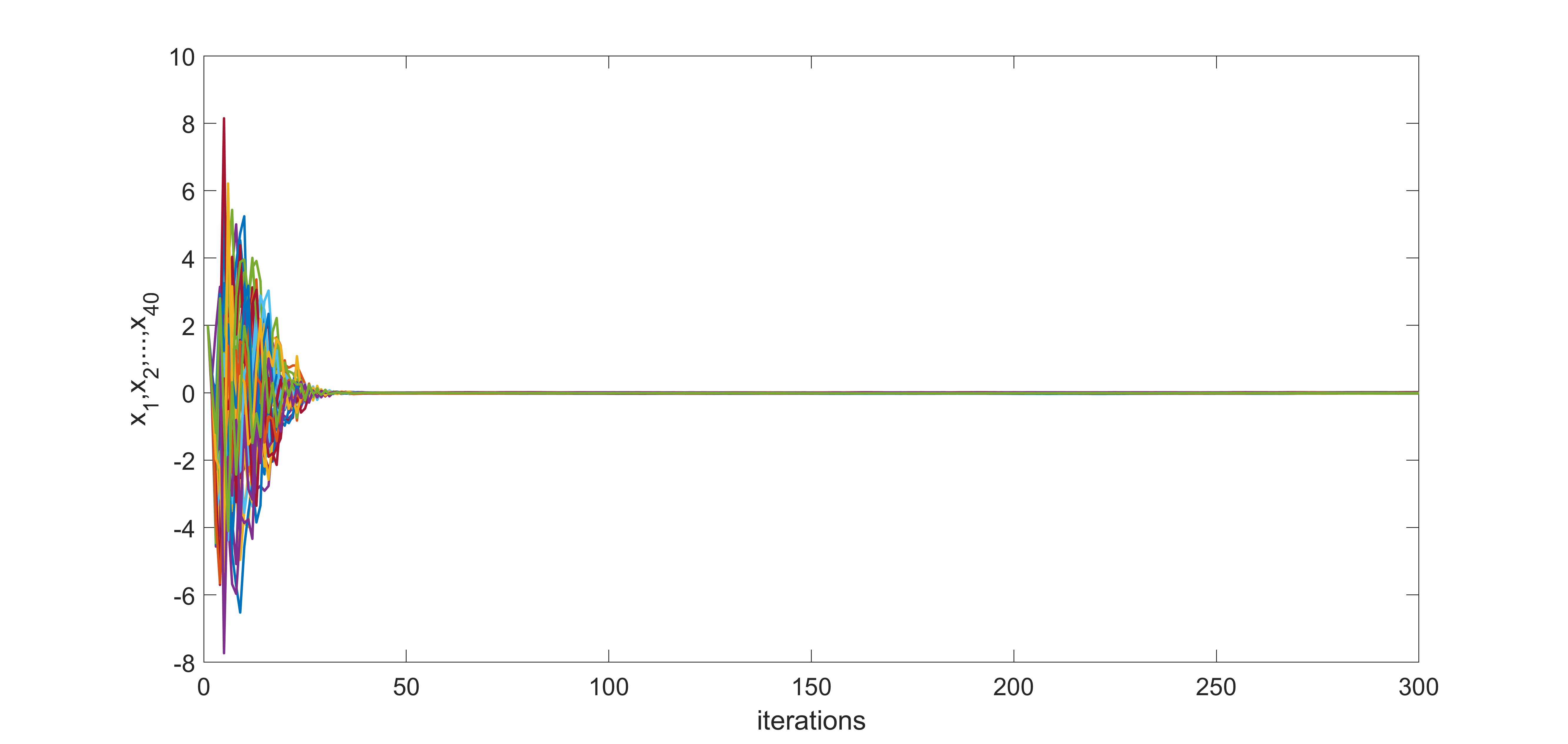}
         \caption{}
         \label{fig:geom:x}
     \end{subfigure}
     \hfill
     \begin{subfigure}[b]{0.45\textwidth}
         \centering
         \includegraphics[width=\textwidth]{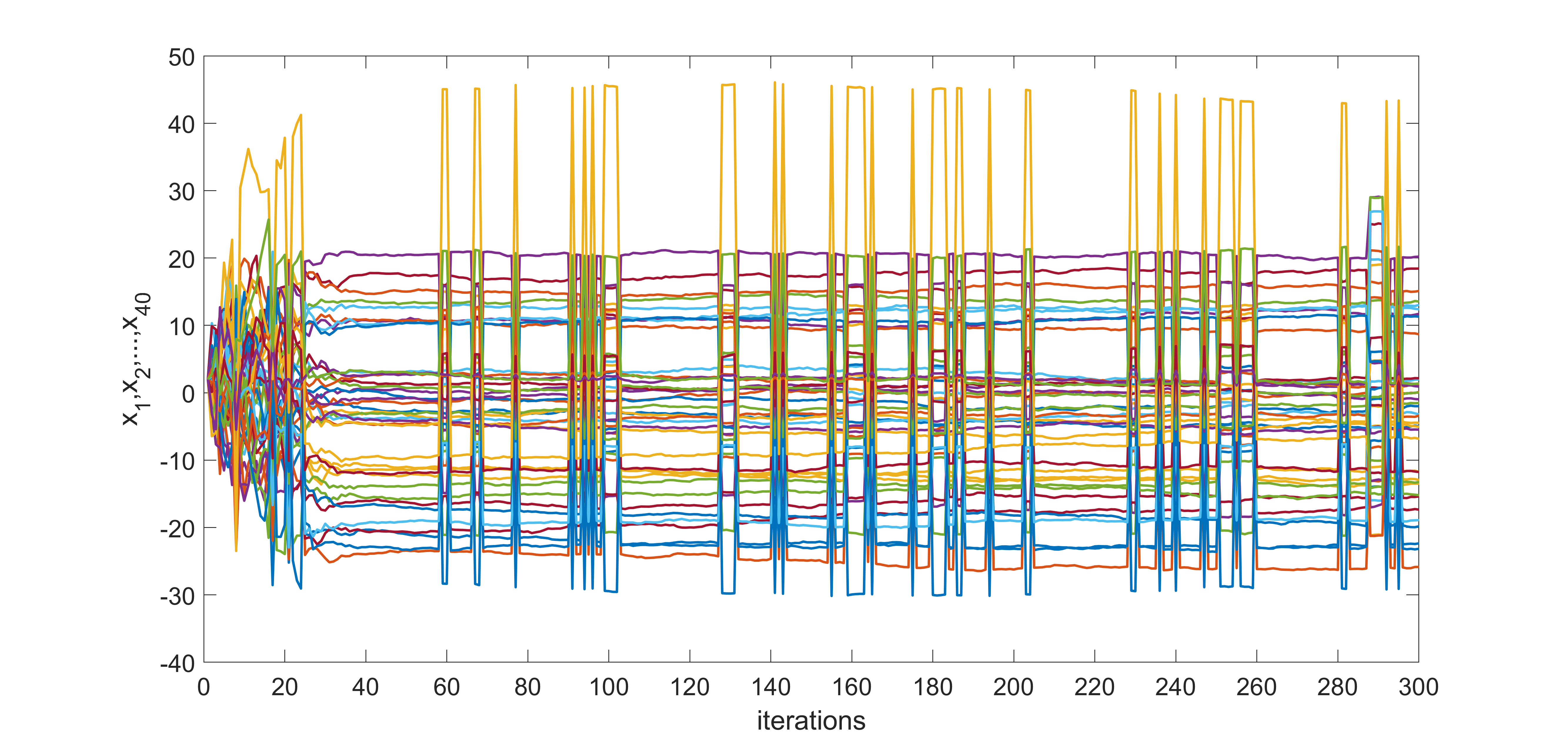}
         \caption{}
         \label{fig:Euclid:x}
     \end{subfigure}
    \hfill
     \begin{subfigure}[b]{0.45\textwidth}
         \centering
         \includegraphics[width=\textwidth]{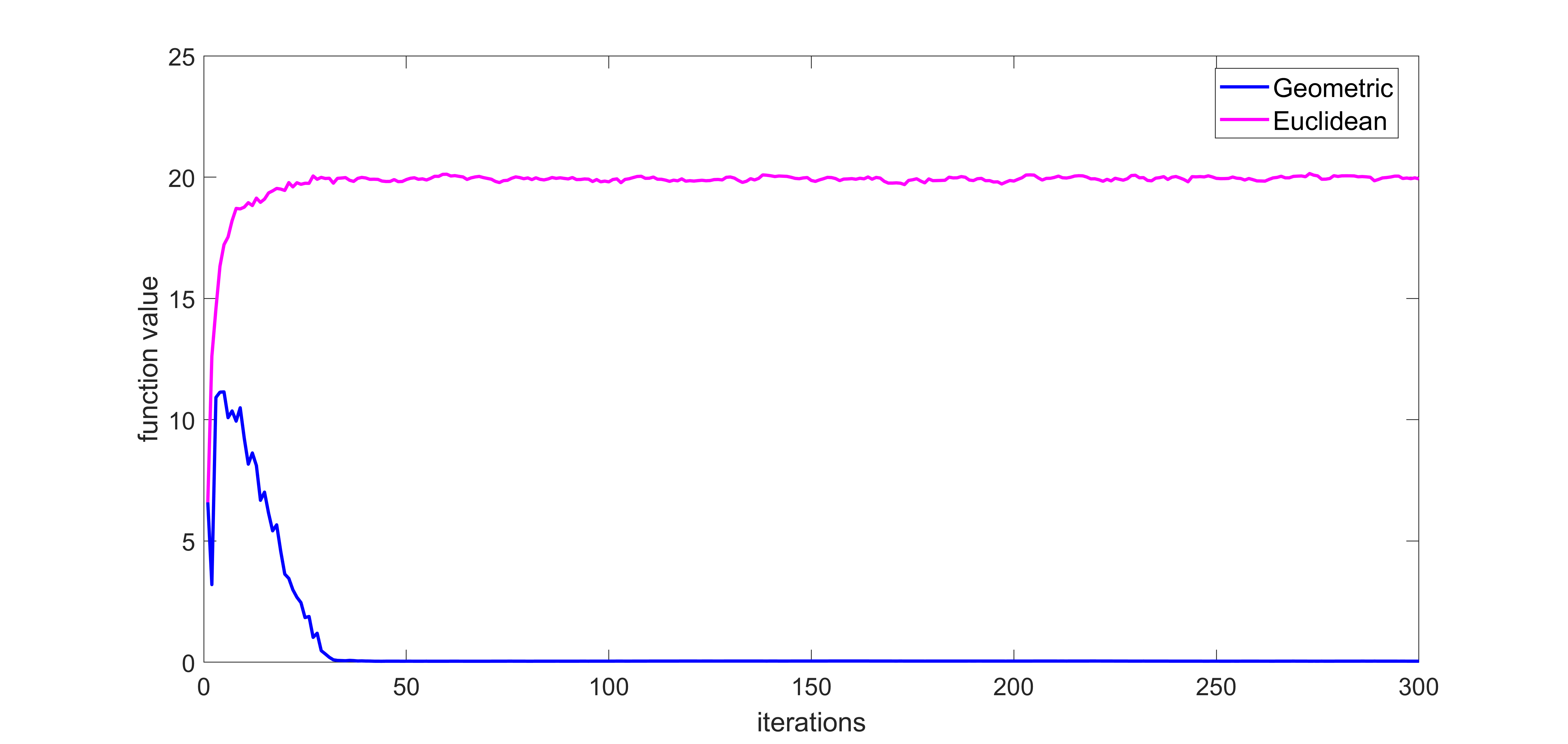}
         \caption{}
         \label{fig:ackley40d:fun}
     \end{subfigure}
        \caption{40-dimensional Ackley function minimization: results by the proposed method are shown in (a) and those via its Euclidean counterpart in (b). Evolutions of function values pertaining to the geometric and Euclidean methods are compared in (c). The problem parameters used are as follows: ensemble size $N=5$,$a=20, b=0.2, c=2 \pi$, $\Delta t = 0.5$, (Euclidean: $\Delta t = 0.0.01$). $\beta$ is an annealing parameter, starting with $\beta_0=50,000$ (Euclidean: $\beta_0=1000$) and reduced with iterations as per $\beta_{k+1} = \beta_k/(\exp{0.01 \times k })$ until $\beta$ becomes less than $0.5$. The regularization parameter $\Upsilon = 10^6$}. 
        \label{fig:Ackley40d}
\end{figure}

\begin{figure}
     \centering
     \begin{subfigure}[b]{0.45\textwidth}
         \centering
         \includegraphics[width=\textwidth]{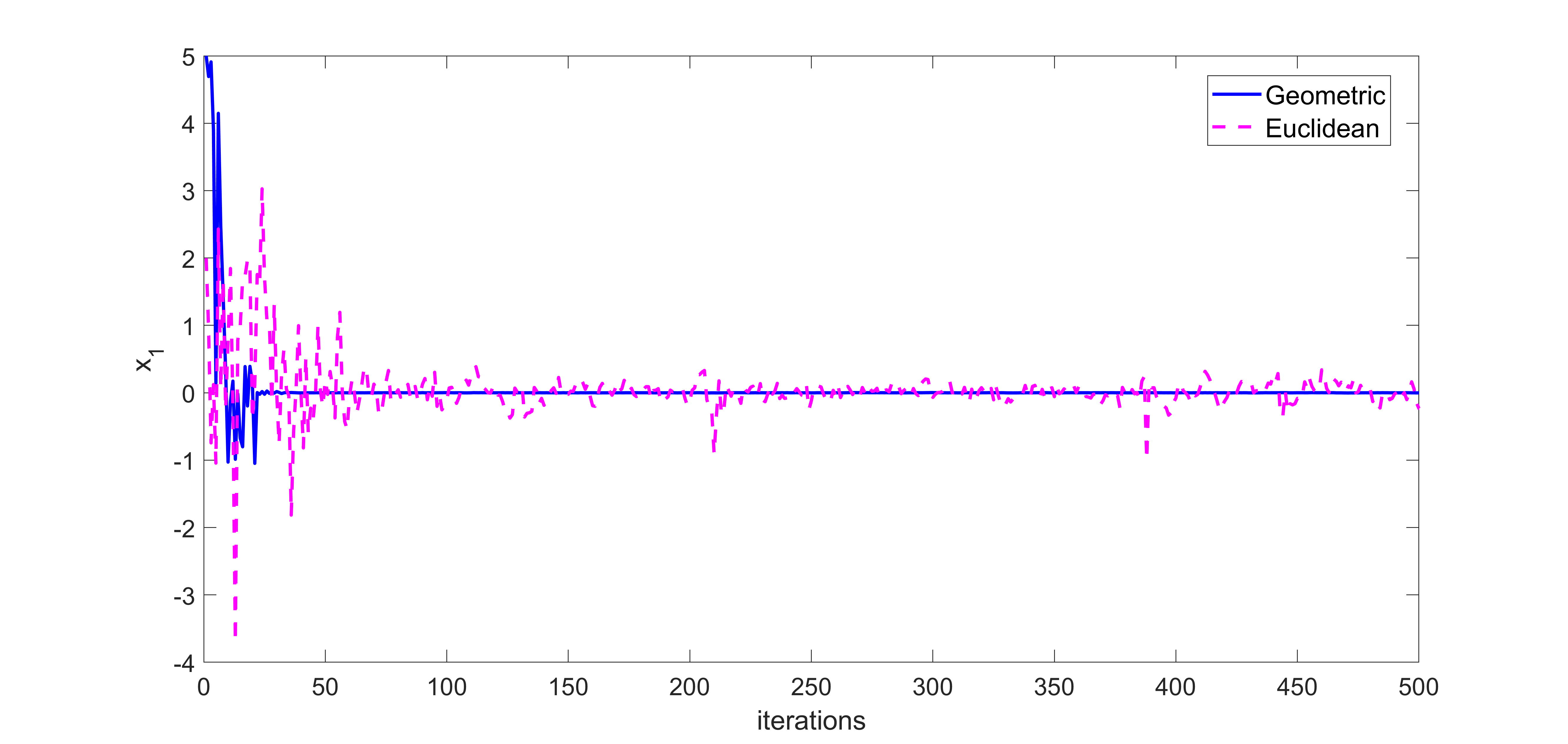}
         \caption{}
         \label{fig:ackley2d:x1}
     \end{subfigure}
     \hfill
     \begin{subfigure}[b]{0.45\textwidth}
         \centering
         \includegraphics[width=\textwidth]{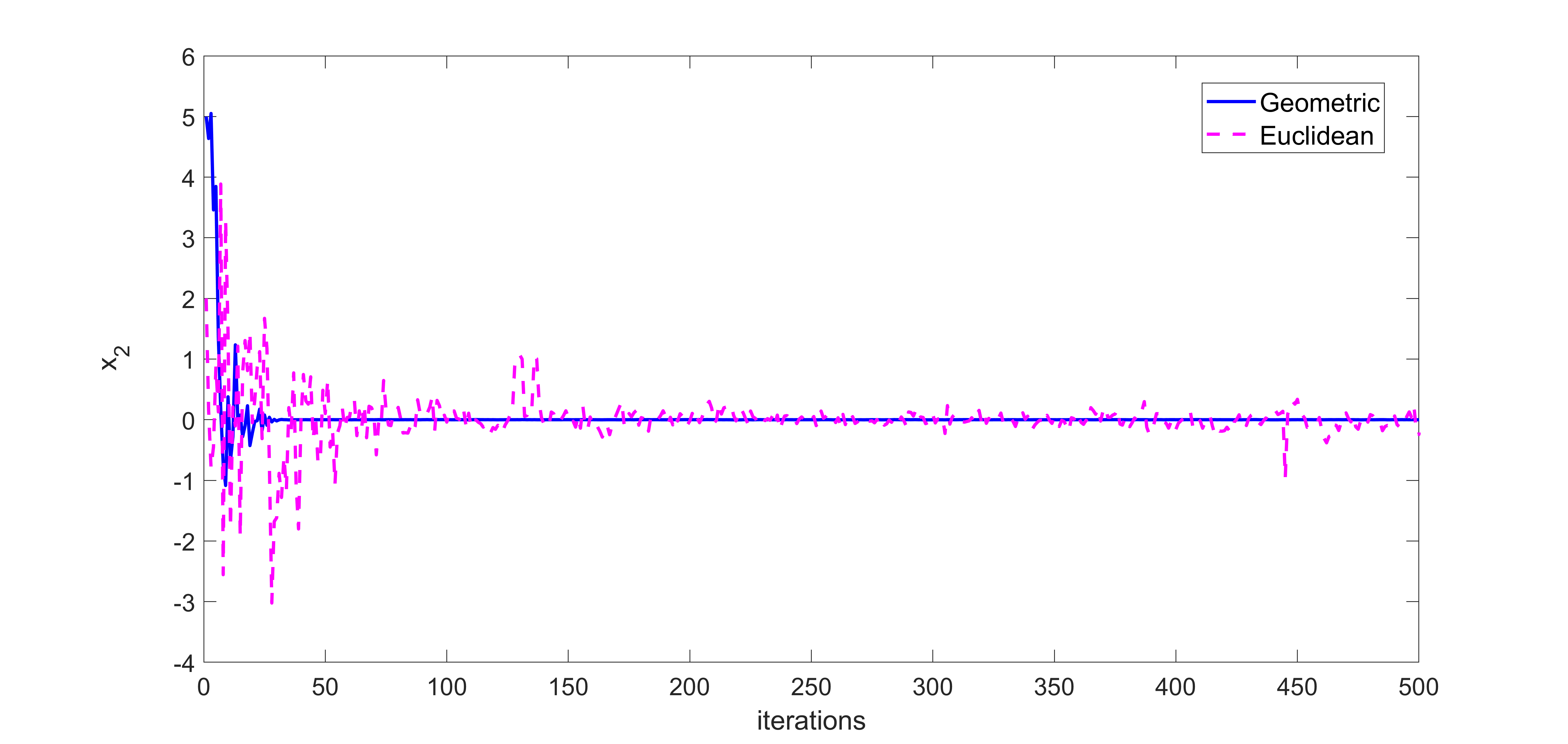}
         \caption{}
         \label{fig:ackley2d:x2}
     \end{subfigure}
    \hfill
     \begin{subfigure}[b]{0.45\textwidth}
         \centering
         \includegraphics[width=\textwidth]{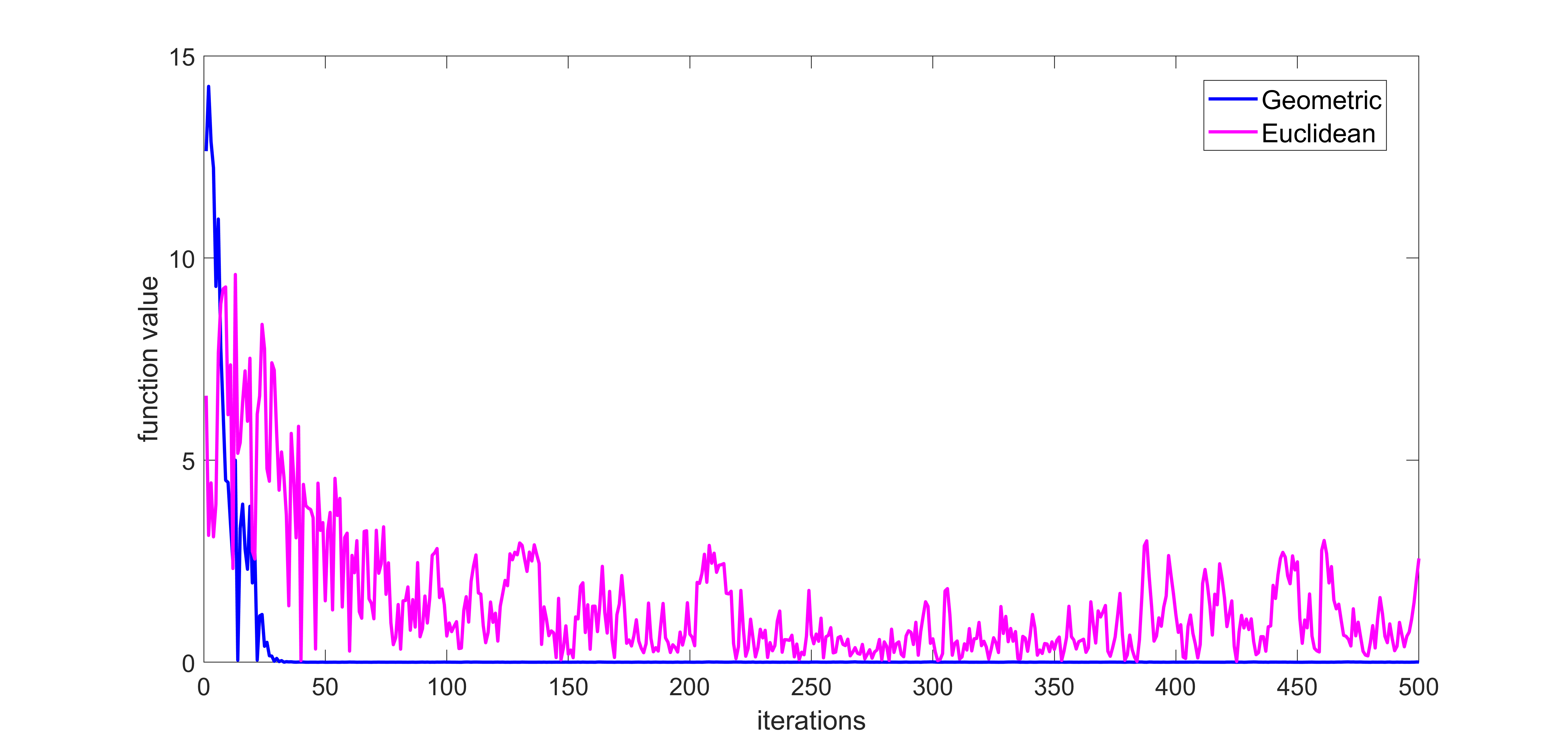}
         \caption{}
         \label{fig:ackley2d:fun}
     \end{subfigure}
        \caption{2-dimensional Ackley function minimization: comparison between the proposed method and its Euclidean counterpart. The $x_1$ and $x_2$ components are shown in subfigures (a) and (b) respectively while the comparison of  function values is shown in (c). Problem parameters are taken as follows: ensemble size $N=5$,$a=20, b=0.2, c=2 \pi$, $\Delta t = 0.5$, (Euclidean: $\Delta t = 0.0.01$). $\beta$ is an annealing like parameter, starting with $\beta_0=1000$ (Euclidean: $\beta_0=50$) and reduced with iterations as per $\beta_{k+1} = \beta_k/(\exp{0.01 \times k })$ until $\beta$ becomes less than $0.5$.The regularization parameter $\Upsilon = 10^6$}.
        \label{fig:Ackley2d}
\end{figure}

The Ackley function $f(x)$ to be minimized is given by:  
\begin{equation}
f(\textbf{x}) = f(x_1, ..., x_n)= -a \exp(-b\sqrt{\frac{1}{n}\sum_{i=1}^{n}x_i^2})-\exp(\frac{1}{n}\sum_{i=1}^{n} \cos(cx_i))+ a + \exp(1)
\end{equation}
where $n$ is the dimension of $x$, the design variable. The overdamped Langevin SDE is of the following form.
\begin{equation}
dX_t= - \beta_t \nabla f(X_t)dt+ \sqrt{ 2 \beta_t} dB_t
\end{equation}
Its stochastically developed version is given by:  
\begin{equation}
dX_t = - \beta_t \sqrt{g^{-1}}\nabla f(X_t)dt +\sqrt{2 \beta_t} \sqrt{g^{-1}}  dB_t - \beta_t g^{-1}\gamma dt
\end{equation}
where 
\begin{equation}
g_{ij}= \frac{1}{2} \frac{\partial^2 f(\textbf{x}) }{\partial x_i \partial x_j} + \Upsilon \delta_{ij}
\end{equation}
The Riemannian connection $\gamma$ can be obtained from the derivatives of $g$ (Appendix C). Since we need to compute $g$ and $\gamma$ to arrive at the developed SDE, our scheme is not gradient-free unlike most others based on metaheuristics, e.g. the genetic algorithm. However, when the gradient of the objective function is available, it is expected that the present approach should have the benefit of a relatively faster convergence. For a 40-dimensional Ackley function, we have reported the results in Figure \ref{fig:Ackley40d}. An ensemble size of only five particles has been used for this purpose. As can be seen in the figure, the solution through the Euclidean route fails to converge for the 40-dimensional problem even as the stochastically developed version converges within 40 steps. We may note that the Euclidean version works for the 2-dimensional case; see Figure \ref{fig:Ackley2d}. However, in this case too, the quality of performance of the geometric version is much better. In reporting these results, the algorithm parameters are so chosen (by trial and error) as to represent  the best performance of each method.

We have only provided the basic outlines of what seems to be a potentially powerful and geometrically inspired stochastic search scheme. While we have adopted an energy-based route for the stochastic search, a geometrically adapted version of a martingale based approach \cite{combeo} could as well be used. Yet another option would be to explore a geometric variant of the stochastic approximation framework \cite{bookkushnerstochastic}. Note that, within the current setup, constrained optimization problems could also be solved through an appropriate modification of the energy, viz. via a penalty term similar to the first two problems. In our future work, we would also be interested in an application of this framework to problems such as stochastic filtering and Markov chain Monte Carlo. 

\section{Concluding remarks}\label{sec:conclusions}
The central theme of this article has been a novel scheme for developing solutions of stochastic differential equations on a Riemannian manifold, leading to a demonstration on how this idea constitutes a powerful tool towards more efficacious numerical solutions for a wide range of problems with applications in science and engineering. The method may be viewed as an extension to a concept well known to mathematicians, viz. a Brownian motion on a Riemannian manifold whose generator is the Laplace-Beltrami operator. Through a range of illustrations -- from the Brownian dynamics in a potential well to the search for the global minimum of a non-convex objective function, we have tried to glean insights into how the stochastically developed solution enforces certain constraints on the flow that are natural to and physically consistent with the underlying dynamics. For instance, by deriving the Riemannian metric using an energy-like barrier, we could readily design a numerical integration scheme that preserved the drift of the mean-energy in a Hamiltonian dynamical system under additive noise. Similarly, by requiring solutions to the overdamped Langevin flow to equilibrate around a minimum of an objective function, we could arrive at a novel stochastic search scheme for non-convex optimization. Indeed, much of the power of the stochastic development method is derived from a good choice of the metric and, as we have shown, the second derivative of energy or some energy-like function could be expediently used for this purpose across a broad spectrum of scenarios. 

It is curious that the Brownian noise, whose development on the manifold results in the only term involving the Levi-Civita connection in the developed equation, should play such a pivotal role in our approach. One wonders if the quality of solutions could be further enhanced by explicitly incorporating information on the Riemannian curvature tensor within the evolving dynamics. This seems feasible if we were to borrow ideas from Cartan's moving frames \cite{ClellandJeanneN} and write the dynamics using the language of exterior calculus whilst exploiting Cartan's structure equations. We wish to take this up in a future article.


\section*{Appendices}
\appendix
\section{Brownian motion in a potential well}
Let $dB_t$ be an $n$-dimensional Brownian motion
\begin{equation} \nonumber
dX_t = dB_t
\end{equation}
The equation for the potential well for $x$ to be near $\beta$ is taken as
\begin{equation}\nonumber
E(x) = \exp{(x-\lambda)^T [\alpha^2] (x-\lambda)}
\end{equation}
Let $[\alpha^2] $ be a diagonal matrix with entries $d_1, d_2...d_n$. This is related to the sharpness of the potential well - higher $d$ values imply a sharper potential well. Then, the energy-like term can be written as
\begin{equation}\nonumber
E(x) = \exp{(d_p(x-\lambda)_p^2  )} 
\end{equation}
Now, 
\begin{eqnarray}
g_{ij} &=& \frac{1}{2} \frac{\partial^2 E(x)}{\partial x_i \partial x_j} \nonumber \\
&=&\frac{1}{2}  \frac{\partial }{\partial x_i}(\frac{\partial E(x)}{\partial x_j}) \nonumber
\end{eqnarray}

\begin{eqnarray}
\frac{\partial E(x)}{\partial x_j} &=& \frac{\partial \exp{(d_p(x-\lambda)_p^2  )}}{\partial x_j} \nonumber \\
&=& 2 d_{(j)}(x-\lambda)_{(j)} \exp{(d_p(x-\lambda)_p^2  )} \nonumber 
\end{eqnarray}
Note that indices in round brackets imply no sum. Therefore, we have
\begin{eqnarray}
\frac{\partial }{\partial x_i}(\frac{\partial E(x)}{\partial x_j}) &=& \frac{\partial \{ 2 d_{(j)}(x-\lambda)_{(j)} \exp{(d_p(x-\lambda)_p^2  )} \}}{\partial x_i}  \nonumber \\
&=& 4 d_{(i)} d_{(j)}(x-\lambda)_{(i)} (x-\lambda)_{(j)}    \exp{(d_p(x-\lambda)_p^2  )} +  2 d_{(j)} \delta_{ij} \exp{(d_p(x-\lambda)_p^2)}  \nonumber
\end{eqnarray}

\begin{eqnarray}
g_{ij} &=& 2 d_{(i)} d_{(j)}(x-\lambda)_{(i)} (x-\lambda)_{(j)}    \exp{(d_p(x-\lambda)_p^2  )} +   d_{(j)} \delta_{ij} \exp{(d_p(x-\lambda)_p^2)} \nonumber
\end{eqnarray}

\textit{Derivatives of $g$}

\begin{eqnarray}
\frac{\partial g_{ij}}{\partial x_k} &=& \frac{\partial \{ 2 d_{(i)} d_{(j)}(x-\lambda)_{(i)} (x-\lambda)_{(j)}    \exp{(d_p(x-\lambda)_p^2  )}\}}{\partial x_k}  + \frac{\partial \{   d_{(j)} \delta_{ij} \exp{(d_p(x-\lambda)_p^2)} \}}{\partial x_k} \nonumber \\
&=&  2 d_{(i)} d_{(j)} (x-\lambda)_{(j)}    \exp{(d_p(x-\lambda)_p^2  )} \delta_{ik}  + 2 d_{(i)} d_{(j)}(x-\lambda)_{(i)}  \exp{(d_p(x-\lambda)_p^2  )}\delta_{jk} \nonumber \\
&& + 4 d_{(i)} d_{(j)}d_{(k)}(x-\lambda)_{(i)} (x-\lambda)_{(j)}     (x-\lambda)_{(k)} \exp{(d_p(x-\lambda)_p^2  )} \nonumber \\
&& +  2 d_{(j)} d_{(k)} \delta_{ij} (x-\lambda)_{(k)} \exp{(d_p(x-\lambda)_p^2  )} \nonumber 
\end{eqnarray}
The equation for the Riemannian connection can be determined from the equations for $g$ and $\frac{\partial g} {\partial x}$.

\section{Drift preserving integration of Duffing equation}

The following equation can be used to determine the mean energy (as a function of time) of a stochastic Hamiltonian system under additive noise \cite{chen2020drift} in terms of its displacement $x_{1,t}$ and velocity $x_{2,t}$.
\begin{equation}\nonumber
Z_t = \mathbb{E}[H(x_{1,t}, x_{2,t})]=\mathbb{E}[H(x_{1,0}, x_{2,0})] + \frac{1}{2} \Tr(\Sigma^T \Sigma) t
\end{equation}
where $\Sigma$ is the noise intensity matrix and $H$ represents the Hamiltonian of the system which for the Duffing oscillator is given by $\frac{x_{2,t}^2}{2}+ \frac{kx_{1,t}^2}{2} + \frac{\alpha x_{1,t}^4}{4}$. Suppressing the time indices from the states, the energy-like term to be used for the drift preserving integration is taken as follows.
\begin{equation}\nonumber
E_t =   \exp{\beta(\frac{x_2^2}{2}+ \frac{kx_1^2}{2} + \frac{\alpha x_1^4}{4} - Z_t)^2} -1
\end{equation}
where $\beta $ is an algorithm parameter. Let 
\begin{equation}\nonumber
V_t = \frac{x_2^2}{2}+ \frac{kx_1^2}{2} + \frac{\alpha x_1^4}{4} 
\end{equation}
Then we have
\begin{equation}\nonumber
E_t = \exp{(\beta [V_t - Z_t]^2)} -1
\end{equation}
The Riemannian metric $g$ can be determined from the energy-like term as follows.
\begin{eqnarray}
g_{ij} &=& \frac{1}{2} \frac{\partial^2 E_t}{\partial x_i \partial x_j} + \Upsilon \delta_{ij}\;,\; i,j={1,2}\nonumber 
\end{eqnarray}
Now,
\begin{eqnarray}
\frac{\partial E_t}{\partial x_j} &=&  2 \beta (V_t-Z_t) \exp{(\beta [V_t - Z_t]^2)}  \frac{\partial V_t}{\partial x_j} \nonumber
\end{eqnarray}
Therefore,
\begin{eqnarray}
\frac{\partial^2 E_t}{\partial x_i \partial x_j} &=& \frac{\partial }{\partial x_i}(\frac{\partial E_t}{\partial x_j}) \nonumber \\
&=& \frac{\partial }{\partial x_i}(  \{2 \beta (V_t-Z_t)  \exp{(\beta [V_t - Z_t]^2)} \} \frac{\partial V_t}{\partial x_j} ) \nonumber \\
&=& 2 \beta \exp{(\beta [V_t - Z_t]^2)} \frac{\partial V_t}{\partial x_j} \frac{\partial V_t}{\partial x_i}  
+\{2\beta (V_t-Z_t)\}^2 \exp{(\beta [V_t - Z_t]^2)} \frac{\partial  V_t  }{\partial x_i}\frac{\partial V_t}{\partial x_j}  \nonumber \\
&&+\{ 2 \beta (V_t-Z_t)  \exp{(\beta [V_t - Z_t]^2)} \} \frac{\partial^2 V_t }{\partial x_i \partial x_j} \nonumber 
\end{eqnarray}

\textit{Derivative of $g$}

The derivative of $g$ along with $g$ itself is required to determine the Levi-Civita or Riemannian connection $\gamma$. We would specifically need the following derivatives.
\begin{equation}\nonumber
\frac{\partial g_{ij}}{\partial x_k} = \frac{1}{2}\frac{\partial^3 E_t}{\partial x_k \partial x_i \partial x_j} 
\end{equation}

\begin{eqnarray}
\frac{\partial^3 E_t}{\partial x_k \partial x_i \partial x_j} &=& \frac{\partial }{\partial x_k}(\frac{\partial^2 E_t}{\partial x_i \partial x_j})  \nonumber \\
&=&\frac{\partial }{\partial x_k}(2 \beta \exp{(\beta [V_t - Z_t]^2)} \frac{\partial V_t}{\partial x_j} \frac{\partial V_t}{\partial x_i} ) \nonumber \\
&&+ \frac{\partial }{\partial x_k}(\{2\beta (V_t-Z_t)\}^2 \exp{(\beta [V_t - Z_t]^2)} \frac{\partial  V_t  }{\partial x_i}\frac{\partial V_t}{\partial x_j} ) \nonumber \\
&&+\frac{\partial }{\partial x_k}(\{1  + 2 \beta (V_t-Z_t)  \exp{(\beta [V_t - Z_t]^2)} \} \frac{\partial^2 V_t }{\partial x_i \partial x_j}) \nonumber 
\end{eqnarray}

For the third order derivatives, let
\begin{equation}\nonumber
A= \frac{\partial }{\partial x_k}(2 \beta \exp{(\beta [V_t - Z_t]^2)} \frac{\partial V_t}{\partial x_j} \frac{\partial V_t}{\partial x_i} ) 
\end{equation}

\begin{equation}\nonumber
 B=  \frac{\partial }{\partial x_k}(\{2\beta (V_t-Z_t)\}^2 \exp{(\beta [V_t - Z_t]^2)} \frac{\partial  V_t  }{\partial x_i}\frac{\partial V_t}{\partial x_j} )
\end{equation}

\begin{equation}\nonumber
C= \frac{\partial }{\partial x_k}(\{1  + 2 \beta (V_t-Z_t)  \exp{(\beta [V_t - Z_t]^2)} \} \frac{\partial^2 V_t }{\partial x_i \partial x_j})
\end{equation}
Then 
\begin{equation}\nonumber
\frac{\partial^3 E_t}{\partial x_k \partial x_i \partial x_j} = A+B+C
\end{equation}
Simplifying $A,B,C$, we have

\begin{eqnarray}
A &=& \{2 \beta \}^2 (V_t-Z_t)\exp{(\beta [V_t - Z_t]^2)} \frac{\partial V_t}{\partial x_k} \frac{\partial V_t}{\partial x_j} \frac{\partial V_t}{\partial x_i}\nonumber \\
&&+ 2 \beta \exp{(\beta [V_t - Z_t]^2)} \frac{\partial V_t}{\partial x_i}   \frac{\partial^2 V_t}{\partial x_k \partial x_j}  \nonumber \\
&&+ 2 \beta \exp{(\beta [V_t - Z_t]^2)} \frac{\partial V_t}{\partial x_j}   \frac{\partial^2 V_t}{\partial x_k \partial x_i}  \nonumber
\end{eqnarray}

\begin{eqnarray}
B&=&8 \beta^2 (V_t-Z_t) \exp{(\beta [V_t - Z_t]^2)} \frac{\partial  V_t  }{\partial x_i}\frac{\partial V_t}{\partial x_j}     \frac{\partial V_t }{\partial x_k} \nonumber \\
&&+\{2 \beta (V_t-Z_t) \}^3 \exp{(\beta [V_t - Z_t]^2)} \frac{\partial  V_t  }{\partial x_i}\frac{\partial V_t}{\partial x_j} 
\frac{\partial V_t}{\partial x_k} \nonumber \\
&&+4 \beta^2 (V_t-Z_t)^2 \exp{(\beta [V_t - Z_t]^2)} \frac{\partial V_t}{\partial x_j}  \frac{\partial^2 V_t}{\partial x_i \partial x_k} \nonumber \\
&&+ 4 \beta^2 (V_t-Z_t)^2 \exp{(\beta [V_t - Z_t]^2)} \frac{\partial  V_t  }{\partial x_i} \frac{\partial^2 V_t }{\partial x_j \partial x_k} \nonumber
\end{eqnarray}

\begin{eqnarray}
C &=&  2 \beta   \exp{(\beta [V_t - Z_t]^2)}\frac{\partial^2 V_t }{\partial x_i \partial x_j} \frac{\partial V_t}{\partial x_k} \nonumber \\
&&+ \{2 \beta (V_t-Z_t)\}^2  \exp{(\beta [V_t - Z_t]^2)} \frac{\partial^2 V_t }{\partial x_i \partial x_j}  \frac{\partial V_t }{\partial x_k} \nonumber \\
&&+\{2 \beta (V_t-Z_t)  \exp{(\beta [V_t - Z_t]^2)} \}  \frac{\partial^3 V_t }{\partial x_k \partial x_i \partial x_j} \nonumber 
\end{eqnarray}

Based on the expressions for $A,B,C$, the derivatives of $g$ can be determined using which along with the expression for $g$, the Levi-Civita connection can be determined.

\section{Non-convex optimization}

\begin{equation}\nonumber
f(\textbf{x}) = f(x_1, ..., x_n)= -a \exp(-b\sqrt{\frac{1}{n}\sum_{i=1}^{n}x_i^2})-\exp(\frac{1}{n}\sum_{i=1}^{n} \cos(cx_i))+ a + \exp(1)
\end{equation}
Treat $f(x)$ as an energy-like function to determine $g$ and $\gamma$. The Langevin SDE to be developed is given by: 
\begin{equation}\nonumber
dX_t= - \beta_t \nabla f(X_t)dt+ \sqrt{ 2 \beta_t} dB_t
\end{equation}
where $\beta$ is an annealing like parameter.
The developed SDE is: 
\begin{equation}\nonumber
dX_t = - \sqrt{g^{-1}}\beta_t \nabla f(X_t)dt + \sqrt{g^{-1}} \beta_t dB_t - \frac{1}{2}g^{-1}\gamma dt
\end{equation}
where
\begin{equation}
g_{ij}= \frac{1}{2} \frac{\partial^2 f(x)}{\partial x_i \partial x_j} +\Upsilon \delta_{ij}
\end{equation}
Let 
$$T_1(x) =  \exp(-b\sqrt{\frac{1}{n}\sum_{i=1}^{n}x_i^2}) $$
$$T_2(x)=  \exp(\frac{1}{n}\sum_{i=1}^{n} \cos(cx_i)) $$
Therefore, $g$ and its derivatives can be written as
\begin{equation}\nonumber
g_{ij} = -\frac{a}{2} \frac{\partial^2 T_1(x)}{\partial x_j \partial x_k} - \frac{1}{2} \frac{\partial^2 T_2(x)}{\partial x_j \partial x_k} 
\end{equation}
and
\begin{equation}\nonumber
\frac{\partial g_{ij}}{\partial x_m} = -\frac{a}{2} \frac{\partial^3 T_1(x)}{\partial x_j \partial x_k \partial x_k} - \frac{1}{2} \frac{\partial^3 T_2(x)}{\partial x_j \partial x_k \partial x_k} 
\end{equation}
We need the first, second and third order derivatives of $T_1,T_2$. \\ 
\textit{First derivative of $T_1$}
\begin{eqnarray}
 \frac{\partial T_1(x)}{\partial x_j}&=&  \frac{\partial  \exp(-\frac{b}{\sqrt{d}} (x_i x_i)^{\frac{1}{2}} ) }{\partial x_j} \nonumber \\
&=& -\frac{b}{\sqrt{d}}  (x_i x_i)^{-\frac{1}{2}}  x_j   T_1(x) \nonumber \end{eqnarray}
\\
\textit{Second derivative of $T_1$}

\begin{eqnarray}
 \frac{\partial^2 T_1(x)}{\partial x_j \partial x_k} &=&   \frac{\partial}{\partial x_k}(-\frac{b}{\sqrt{d}}  (x_i x_i)^{-\frac{1}{2}}  x_j   T_1(x)) \nonumber \\
  &=& \frac{b}{\sqrt{d}}  x_j x_k   (x_i x_i)^{-\frac{3}{2}}T_1(x) -\frac{b}{\sqrt{d}}  (x_i x_i)^{-\frac{1}{2}}  \delta_{jk}  T_1(x) -\frac{b}{\sqrt{d}} (x_i x_i)^{-\frac{1}{2}}  x_j   \frac{\partial (  T_1(x))}{\partial x_k} \nonumber
\end{eqnarray}
\\
\textit{Third derivative of $T_1$}

\begin{eqnarray}
\frac{\partial^3 T_1(x)}{\partial x_j \partial x_k \partial x_m}&=& \frac{b}{\sqrt{d}}\frac{\partial}{\partial x_m}(  x_j x_k   (x_i x_i)^{-\frac{3}{2}}T_1(x) )  - \frac{b}{\sqrt{d}} \frac{\partial}{\partial x_m}(   (x_i x_i)^{-\frac{1}{2}}  \delta_{jk}  T_1(x))  \nonumber \\
&&- \frac{b}{\sqrt{d}} \frac{\partial}{\partial x_m}(  (x_i x_i)^{-\frac{1}{2}}  x_j   \frac{\partial (  T_1(x))}{\partial x_k}) \nonumber \\[15pt]
&=& \frac{b}{\sqrt{d}}    x_k   (x_i x_i)^{-\frac{3}{2}}T_1(x)  \delta_{jm} + \frac{b}{\sqrt{d}}  x_j   (x_i x_i)^{-\frac{3}{2}}T_1(x)  \delta_{km} \nonumber \\[15pt]
&& -3 \frac{b}{\sqrt{d}}   x_j x_k x_m    (x_i x_i)^{-\frac{5}{2}} T_1(x) + \frac{b}{\sqrt{d}}  x_j x_k   (x_i x_i)^{-\frac{3}{2}}  \frac{\partial T_1(x)}{\partial x_m}\nonumber \\[15pt]
&&  \frac{b}{\sqrt{d}}      \delta_{jk} x_m (x_i x_i)^{-\frac{3}{2}} T_1(x)  - \frac{b}{\sqrt{d}}   (x_i x_i)^{-\frac{1}{2}}  \delta_{jk}   \frac{\partial T_1(x)}{\partial x_m} \nonumber \\[15pt]
&& + \frac{b}{\sqrt{d}}    x_j  x_m(x_i x_i)^{-\frac{3}{2}}  \frac{\partial   T_1(x)}{\partial x_k} 
 - \frac{b}{\sqrt{d}}   (x_i x_i)^{-\frac{1}{2}}     \frac{\partial   T_1(x)}{\partial x_k} \delta_{jm}
 - \frac{b}{\sqrt{d}}  x_j   (x_i x_i)^{-\frac{1}{2}}   \frac{\partial^2  T_1(x)}{\partial x_k \partial x_m  } \nonumber
\end{eqnarray}

\textit{First derivative of $T_2$}

\begin{eqnarray}
 \frac{\partial T_2(x)}{\partial x_j} &=&  \frac{\partial }{\partial x_j} (\exp{(\frac{1}{d} \sum^d_{i=1}{\cos{(cx_i)}} )}) \\
 &=& - \frac{c}{d}  \sin{(cx_j)} T_2(x)
\end{eqnarray}

\textit{Second derivative of $T_2$}

\begin{eqnarray}
 \frac{\partial^2 T_2(x)}{\partial x_j \partial x_k}  &=& - \frac{c}{d} \frac{\partial }{\partial x_k} ( \sin{(cx_j)} T_2(x)) \nonumber \\
 &=&  - \frac{c^2}{d} \delta_{jk}   \cos{(cx_j)} T_2(x) - \frac{c}{d} \sin{(cx_j)} \frac{\partial  T_2(x)}{\partial x_k}  \nonumber 
\end{eqnarray}

\textit{Third derivative of $T_2$}

\begin{eqnarray}
\frac{\partial^3 T_2(x)}{ \partial x_m \partial x_j \partial x_k } &=& \frac{\partial}{\partial x_m} ( - \frac{c^2}{d} \delta_{jk}   \cos{(cx_j)} T_2(x) - \frac{c}{d} \sin{(cx_j)} \frac{\partial  T_2(x)}{\partial x_k}  ) \nonumber  \\
&=&  \frac{c^3}{d} \delta_{km} T_2(x) \sin{(cx_j)} - \frac{c}{d}\sin{(cx_j)}  \frac{\partial^2  T_2(x)}{\partial x_k \partial x_m} \nonumber  \\
&& - \frac{c^2}{d} \delta_{jk}\cos{(cx_j)} \frac{\partial T_2(x)}{\partial x_m}   -\frac{c^2}{d}\delta_{jm} \frac{\partial  T_2(x)}{\partial x_k} ( \cos{(cx_j)}) 
\end{eqnarray}

\bibliographystyle{abbrv}
\bibliography{references}

\end{document}